\documentclass[aps,twocolumn,showpacs,superscriptaddress]{revtex4}

\usepackage{graphicx}
\usepackage{amsfonts}
\usepackage{amssymb}
\usepackage{amsbsy}
\usepackage{amsmath}
\usepackage{mathrsfs}
\usepackage{latexsym}
\usepackage{natbib}
\usepackage{bm}
\usepackage{subfigure} 
\usepackage{color}
\usepackage{wasysym}
\usepackage{mathbbol}
\usepackage{bigints}
\allowdisplaybreaks

\def\k{\mathrm{k}}
\def\kr{\kappa}

\def\ksg{\mathrm{\varkappa}}

\def\Phit{\tilde{\Phi}}
\def\omegat{\tilde{\omega}}

\def\rs{r_s}
\def\rstar{r_{\star}}

\def\rhash{r_{\sharp}}
\def\scriplus{\mathscr{I}^{+}}
\def\scriminus{\mathscr{I}^{-}}

\def\observerminus{\mathbb{O}^{-}}
\def\observerplus{\mathbb{O}^{+}}

\def\mstar{m_{\star}}

\begin{document}

\title{Hawking effect in an extremal Kerr black hole spacetime}

\author{Saumya Ghosh}
\email{sg14ip041@iiserkol.ac.in}
\affiliation{Department of Physical Sciences, Indian Institute of Science Education and Research Kolkata, Mohanpur - 741246, WB, India}

\author{Subhajit Barman}
\email{subhajit.b@iitg.ac.in}
\affiliation{Department of Physics, Indian Institute of Technology Guwahati, Guwahati - 781039, Assam, India}


\date{\today}

\begin{abstract}

It is well known that extremal black holes do not Hawking radiate, which is usually realized by 
taking an extremal limit from the nonextremal case. However, one cannot perceive the same phenomenon 
using the Bogoliubov transformation method starting from an extremal black hole itself, 
\textit{i.e.}, without the limiting case consideration. In that case, the Bogoliubov coefficients do 
not satisfy the required normalization condition. In canonical formulation, which closely mimics the 
Bogoliubov transformation method, one can consistently reproduce the vanishing number density of 
Hawking quanta for an extremal Kerr black hole. In this method, the relation between the spatial 
near-null coordinates, imperative in understanding the Hawking effect, was approximated into a sum 
of linear and inverse terms only. In the present work, we first show that one can reach the same 
conclusion in canonical formulation even with the complete relationship between the near-null 
coordinates, which contains an additional logarithmic term. It is worth mentioning that in the 
nonextremal case, a similar logarithmic term alone leads to the thermal Hawking radiation. 
Secondly, we study the case with only the inverse term in the relation (i.e., when the spatial 
near-null coordinates associated to the past and future observers are inversely related to each 
other) to understand whether it is the main contributing term in vanishing number density.
Third, for a qualitative realization, we consider a 
simple thought experiment to understand the corresponding Hawking temperature and conclude that the 
inverse term indeed plays a crucial role in the vanishing number density.

\end{abstract}

\maketitle

\section{Introduction}\label{Introduction}

The Hawking effect remains to be one of the most pioneering results perceived through the use of 
quantum field theory in a black hole spacetime. In \cite{hawking1975} Hawking showed that an 
asymptotic future observer in a black hole spacetime observes a thermal distribution of particles 
which mathematically was realized through the Bogoliubov transformation between the ingoing and 
outgoing field modes. These field modes are again described in terms of the null coordinates that 
must satisfy a logarithmic relation among themselves to discern the Planckian distribution of the 
Hawking quanta \cite{hawking1975}. While this Bogoliubov transformation method is one of the most 
straightforward methods to realize the Hawking effect, it is unable to provide satisfactory results 
in a few areas. One encounters one of such cases while dealing with an extremal black hole. In this 
regard, it is known that the extremal black holes do not Hawking radiate \cite{Barbachoux:2002tt, 
Angheben:2005rm, Vanzo:2011wq}, which is noticed by taking the extremal limit from the nonextremal 
case and also using procedures like the tunneling formulation, Euclidean path integral formalism 
\cite{Angheben:2005rm, Hayward:2008jq, Vanzo:2011wq, Anderson:1995fw, BelgiornoF, Moretti:1995fu, 
Eskin:2019nbo, Parikh:1999mf, PhysRevD.15.2752}. However, in the Bogoliubov transformation method, 
starting from an extremal black hole, the coefficients do not satisfy the necessary consistency 
condition emerging from the commutator of the ladder operators of the field modes 
\cite{Alvarenga:2003tx}. It debars one to obtain the number density of the Hawking quanta reliably.

The authors in \cite{Barman:2017fzh} provided a Hamiltonian-based derivation of the Hawking effect 
in a static Schwarzschild black hole background. One of the primary difficulties implementing a 
Hamiltonian-based framework to realize the Hawking effect was the null coordinates that describe 
the field modes related to the Bogoliubov transformation. These null coordinates do not lead to a 
true matter field Hamiltonian that can provide the evolution of the field modes. In 
\cite{Barman:2017fzh}, a set of near-null coordinates with spacelike and timelike signatures was 
introduced to circumvent this issue and construct the field Hamiltonians.
The canonical formulation \cite{Barman:2017fzh, Barman:2017vqx, Barman:2018ina} closely mimics the 
Bogoliubov transformation method and utilizing it one can consistently reproduce the vanishing 
number density of the Hawking quanta in an extremal Kerr black hole spacetime \cite{Barman:2018ina}. 
Using this Hamiltonian formulation, the Hawking effect in a Schwarzschild and nonextremal Kerr black 
hole spacetime was realized in \cite{Barman:2017fzh, Barman:2017vqx, Barman:2018ina}. 
In these works the relation between the spatial near-null coordinates of two asymptotic observers 
plays a crucial role in the realization of Hawking effect.
In \cite{Barman:2018ina} the authors established an exact relation between the spatial near-null 
coordinates but they used an approximated part of the complete relation to arrive at the vanishing 
number density of the Hawking quanta in a consistent fashion. This approximated relation contains 
one linear and one inverse term. However, in the literature using Bogoliubov transformation method 
only a similar inverse term in the relation between the null coordinates were taken into account  
for the estimations \cite{Liberati:2000sq, Alvarenga:2003tx}. This led to an inconsistency in 
satisfying the normalization condition between the Bogoluibov coefficients. Therefore, it inspires 
us to perform a study with the complete relation between the spatial near-null coordinates in the 
context of canonical formulation. It is important to reconfirm and solidify the claims of previous 
results, also to find out which part of the relation is truly responsible for the vanishing number 
density of the Hawking quanta.

In this work, we consider the exact and complete relation between the near-null coordinates in an 
extremal Kerr black hole spacetime. A detailed study with the complete relation is important in its own right. This relation is a sum of linear, logarithmic and inverse 
functions. The logarithmic term is important since such a similar term leads to the thermal spectrum of 
Hawking radiation in case of non-extremal Kerr black holes. We use the canonical formulation to 
obtain the number density of the Hawking quanta. We arrive at the conclusion that this entire 
general relation is also capable of providing the same conclusion of vanishing number density 
consistently. Furthermore, we also consider only the inverse relation between the spatial near-null 
coordinates, as this brings a close relevance to most of the works in the literature, and observe 
that even in this case mathematically, one can consistently get the vanishing number density. We 
also point out the subtleties and conceptual barriers to describe Hawking effect with this inverse 
relation approximation. Finally, we present a proper understanding of the primary contributing term 
in the vanishing number density by a simple thought experiment. Here outgoing particles nearly 
escaping being trapped by the event horizon in a nonextremal Kerr black hole spacetime reaches an 
asymptotic future observer with a wavelength inversely proportional to the temperature of the 
Hawking effect, which is the \textit{Wien's displacement law} for thermal distribution. On the other 
hand, for an extremal Kerr black hole, this wavelength tends to infinity, thus suggesting that the 
temperature corresponding to Hawking effect is zero. Moreover, we notice that the inverse term in 
the relation between the near-null coordinates is the main contributing term in this vanishing 
temperature of the Hawking effect in an extremal Kerr black hole spacetime.

In Sec. \ref{Sec:Kerr_BH_spacetime}, we provide a brief introduction about the Kerr black hole spacetime. In particular, in this section we talk about the horizon structure and the condition for extremality and set up the background for studying a massless, minimally coupled, free scalar field in this spacetime. In the succeeding Sec. \ref{Sec:canonical-framework} we present an overview of the canonical formulation with the near-null coordinates. In Sec. \ref{Sec:extremal-kerr-Hawking} we specifically consider the extremal situation in the Kerr black hole spacetime and, using the canonical formulation, estimate the consistency condition and the number density of the Hawking quanta. We mention that in different sub-sections of this particular section, we shall be estimating the consistency and the number density using disparate relevant relations among the near-null coordinates. Subsequently, in Sec. \ref{Sec:wiens-law} we prepare a physically understandable set-up to shed further light on the dominating term in the relation between the near-null coordinates, which contributes to the vanishing number density. We conclude with a discussion of our findings in Sec.\ref{Sec:discussion}.

In this work we consider the natural units, i.e., the speed of light in vacuum $c$ and Planck constant $\hbar$ will have unit values.

\section{HorizonS, Extremality and scalar field in a Kerr black hole 
spacetime}\label{Sec:Kerr_BH_spacetime}

In this section, first, we are going to give a brief overview of the Kerr black hole spacetime. In 
particular, we will represent its metric in terms of the Boyer-Lindquist coordinates, elucidate the 
position of the two horizons, and the condition of extremality when the two horizons merge to a 
single one. Second, we will discuss the characteristics of a massless minimally coupled scalar field 
in this spacetime. Specifically, we shall talk about the action of this scalar field in regions near 
the horizon and radial infinity. The purpose is to show that one can utilize the understandings of 
quantum field theory from the flat spacetime in these regions and realize the Hawking effect.

\subsection{The Kerr black hole spacetime}
The Kerr black hole spacetime represents an exact solution of the vacuum 
Einstein field equations outside of a rotating mass. Unlike the hypothetical 
charged (Reissner-Nordstr\"om) and charged rotating (Kerr-Newman) black holes, 
the Kerr black holes have gained immense astrophysical significance, especially 
after the detection of the gravitational waves \cite{Abbott:2016blz, 
Abbott:2016nmj, TheLIGOScientific:2016pea, Abbott:2017vtc} from the perceived 
merger of these rotating black holes. In particular, the mass $M$ and angular 
momentum per unit mass $a$ are the sole delineating parameters which describe a 
Kerr black hole. One can express the line element in this spacetime Using the 
Boyer-Lindquist coordinates \cite{Boyer:1967} as 
\begin{eqnarray}\label{eq:Kerr-Metric-in-Boyer-Lindquist}
ds^2 = -  \frac{1}{\rho^2} (\Delta -a^2\sin^2{\theta}) dt^2 + 
\frac{\rho^2}{\Delta} dr^2 + \rho^2 d\theta^2   \nonumber\\
~ + \frac{\Sigma}{\rho^2} \sin^2{\theta}~d\phi^2 - 
\frac{2a}{\rho^2} (r^2+a^2-\Delta) \sin^2{\theta} ~dt d\phi ~, 
\end{eqnarray}
where $\rho^2=r^2+a^2\cos^2{\theta}$, $\Sigma=(r^2+a^2)^2-a^2 
\Delta\sin^2{\theta}$, $\Delta=r^2+a^2-r_{s}r$, and $r_{s}=2GM$ with $G$ 
representing the Newton's gravitational constant \cite{BookCarroll, Kerr2007dk, 
BookPoisson, Dadhich2013qx, BookSchutz, BookRaine, BookPadmanabhanGrav, 
Heinicke2015iva, Krasinski1900zzb, Teukolsky2014vca, Visser2007fj, 
Smailagic2010nv}. It should be mentioned that in this spacetime one obtains the 
coordinate singularity for $\Delta=0$ and the curvature singularity for 
$\rho^2=0$. In the latter case the \textit{Kretschmann scalar} is singular and 
mere coordinate transformations cannot remove this type of singularity. On the 
other hand, from the condition of the coordinate singularity $\Delta=0$ one can 
find out the positions of the apparent horizons $r=r_{h}$ and $r=r_{c}$ as 
\begin{equation}\label{eq:horizons}
r_h= \tfrac{1}{2}(r_s + \sqrt{r_s^2 - 4 a^2} ~) ~,~
r_c= \tfrac{1}{2}(r_s - \sqrt{r_s^2 - 4 a^2} ~) ~.
\end{equation}
Here $r_h$ and $r_c$ represent the outer event horizon and the Cauchy horizon, with 
$\ksg_h=\sqrt{r_s^2-4a^2}/(2r_s r_h)$ and $\ksg_c=\sqrt{r_s^2-4a^2}/(2r_s r_c)$ being their 
respective surface gravities. An interesting phenomenon in a Kerr black hole spacetime is an 
inertial observer is not static due to the frame-dragging effect (for example see chapter 5, page 
188 of \cite{BookPoisson}, and chapter 11, page 310 of \cite{BookSchutz}) and experiences an angular 
velocity
\begin{equation}\label{InertialAngularVelocity}
\Omega \equiv \Omega(r,\theta) = \frac{g^{t\phi}}{g^{tt}} 
= \frac{a r \rs}{\Sigma}   ~.
\end{equation}
The effect of this frame dragging results in a non zero $\dot{\phi} = a/\Delta$ 
in the null geodesics' governing equations \cite{BookPoisson}, which is unlike 
the case in the Schwarzschild spacetime. The other governing equations for null 
geodesics here are $\dot{t} = (r^2+a^2)/\Delta,~\dot{r} = \pm 1 ~,~ \dot{\theta} 
= 0$, where the overhead dot denotes derivative with respect to some affine 
parameter. Using these equations one can perceive that along the ingoing null 
trajectories, the coordinates $v=t+\rstar$ and $\psi=\phi+\rhash$ are constants, 
while along the outgoing null trajectories the coordinates $u=t-\rstar$ and 
$\chi=\phi-\rhash$ are constants. The expressions of $\rstar$ (the 
\textit{tortoise} coordinate) and $\rhash$ are obtained from 
\begin{equation}\label{eq:diff-tortoise}
d\rstar=\frac{r^2+a^2}{\Delta}dr ~~,~~ d\rhash=\frac{a}{\Delta}dr~.
\end{equation}
As we will study the Hawking effect in an extremal Kerr black hole spacetime, it 
is imperative to talk about the $\rstar$ in this case. In the extremal case, 
both the horizons from Eq. (\ref{eq:horizons}) merge together which corresponds 
to $a\to r_{s}/2$. In this case, from the definition (\ref{eq:diff-tortoise}) 
one can find out the expressions of the coordinates $\rstar$ and $\rhash$ for an 
extremal Kerr black hole as 
\begin{equation}\label{eq:tortoise-extremal}
\rstar=r+r_{s} \ln 
\left(\frac{2r - r_s }{r_s}\right)-\frac{r_s^2}{2r-r_s}  ~;~~\rhash=-\frac{2a}{2r-r_s}~.
\end{equation}
These expressions of $\rstar$ and $\rhash$ in terms of radial coordinate $r$ 
differ depending on the extremal or non-extremal case. One also finds the 
tortoise coordinate to be singular \cite{Liberati:2000sq, Alvarenga:2003tx, 
Angheben:2005rm, Gao:2002kz} if one takes the limit $a\to r_{s}/2$ from the 
non-extremal case to get this expression.

\subsection{Scalar field action in a Kerr black hole spacetime}

To understand the consequences of semi-classical gravity in an extremal Kerr black hole spacetime we first consider a massless, minimally coupled free scalar field $\Phi(x)$ in a general spacetime background. The action of scalar field $\Phi(x)$ is given by
\begin{equation}\label{eq:Scalar_field_action}
S_{\Phi} = \int d^{4}x \left[ -\frac{1}{2} \sqrt{-g} g^{\bar{\mu} \nu} \nabla_{\bar{\mu}}\Phi(x) \nabla_{\nu}\Phi(x) \right] ~.
\end{equation}
From \cite{Barman:2018ina} it is seen that this $(1+3)$ dimensional general action (\ref{eq:Scalar_field_action}) can be transformed into a simple $(1+1)$ dimensional form in Kerr black hole spacetime in the regions near the event horizon and near scriplus and scriminus. For this purpose one can consider that due to the axial symmetry of the Kerr spacetime the scalar field can be decomposed as
\begin{eqnarray}
\Phi(t,r,\theta,\phi) = \sum_{lm} e^{i m\phi}~\Phi_{lm}(t,r,\theta)~.
\end{eqnarray}
Putting this decomposition back into action (\ref{eq:Scalar_field_action}) and integrating out the action over the azimuth angle $\phi$ one can further consider a redefinition of the field as
\begin{eqnarray}\label{eq:Rel-PhysicalRedefiben-field}
\Phi_{lm}(t,r,\theta) \equiv e^{-i m \Omega t}~\Phit_{lm}(t,r,\theta)~.
\end{eqnarray}
This will remove all the quantities with a single term of derivative with respect to time, i.e., terms with single $\partial_{t}\Phi_{lm}$. Then considering $\Phit_{lm}(t,r,\theta) = \mathscr{S}_{lm}(\theta)~ \varphi_{lm} (\rstar,t)/\sqrt{r^2+a^2}~,$ and using the spheriodal harmonics orthogonality condition $\int d(\cos{\theta}) \mathscr{S}_{lm} 
(\theta) \mathscr{S}^*_{l'm}(\theta) = \delta_{l,l'}$ one can reduce action $S_{\Phi}$ near horizon as well as near past and future null infinities as $S_{\Phi} = \sum_{lm} S_{lm}$. Here $S_{lm}$ is given by
\begin{equation}\label{eq:2d-reduced-scalar-action}
S_{lm} \simeq \int dt d\rstar 
\left[\tfrac{1}{2}
\partial_t \varphi^*_{lm}\partial_t\varphi_{lm} 
- \tfrac{1}{2}
\partial_{\rstar}\varphi^*_{lm}\partial_{\rstar}\varphi_{lm}
\right] ~,
\end{equation}
which represents a scalar field action in $(1+1)$ dimensional flat Minkowski spacetime.
We mention that with partial field decomposition for the angular coordinates, the scalar field 
action in a Kerr black hole spacetime is realized as an infinite collection of $(1+1)$ dimensional 
fields. This fact is well understood in the literature (for example, see the derivation of Eq. 
(\ref{eq:Scalar_field_action}) in \cite{Iso:2006ut} and the discussions therein). As we have done in 
our work, one can further simplify the action by redefining the field, with relation like Eq. 
(\ref{eq:Rel-PhysicalRedefiben-field}) and introducing a factor of $1/\sqrt{r^2+a^2}$, see also 
\cite{Barman:2018ina}. Then in asymptotically large $r$ and near the event horizon, the field 
imitates the  $(1+1)$ dimensional flat spacetime like shown in Eq. 
(\ref{eq:2d-reduced-scalar-action}) with coordinates $t$ and $r_{\star}$.
It should be mentioned that for the Hawking effect the modes essential are constructed near the 
horizon. In this region if one assigns a frequency $\omegat$ to the redefined field 
$\Phit_{lm}(t,r,\theta)$, i.e., $\Phit_{lm}(t,r,\theta)\sim e^{-i\omegat t}$, then the physical 
field $\Phi_{lm}$ will have the frequency $\omega = \omegat + m \Omega_{h}$. This fact can be 
realized from Eq. (\ref{eq:Rel-PhysicalRedefiben-field}), with the identification of 
$\Omega=\Omega_{h}$ at the horizon, see \cite{Ford:1975tp, Hod:2011zzd, Menezes:2016SeaKs, 
Menezes:2017oeb, Lin:2009wm, Miao:2011dy, Iso:2006ut}. Then for our subsequent discussions regarding 
the Hawking effect in an extremal Kerr black hole spacetime we shall consider the particular action 
(\ref{eq:2d-reduced-scalar-action}) with the perceived transformation to the frequency $\omega$ of 
the physical field as
\begin{equation}\label{eq:FrequencyShift}
\omegat = \omega - m\Omega_h ~. 
\end{equation}
To briefly outline the issues, we mention that the number density of the Hawking quanta in a 
nonextremal Kerr black hole spacetime and as seen by an asymptotic future observer (for a detailed 
analysis, see \cite{Barman:2018ina}) is given by
\begin{eqnarray}\label{eq:ND-nonextremal}
N_{\tilde{\omega}} = \frac{1}{e^{2\pi\tilde{\omega}/\ksg_{h}}-1} =  
\frac{1}{e^{2\pi(\omega-m\Omega_{h})/\ksg_{h}}-1}~,
\end{eqnarray}
where, as mentioned earlier, $\ksg_{h}$ denotes the surface gravity at the event horizon. From 
this spectrum of particles, one may realize the characteristic temperature of the Hawking effect to 
be
\begin{eqnarray}
T_{H} = \ksg_{h}/(2\pi\,k_{B}) = \sqrt{r_{s}^2-4a^2}/(4\pi\,k_{B}r_{s}r_{h})~,
\end{eqnarray}
where $k_{B}$ denotes the Boltzmann constant. Now, one can observe that in the extremal limit $a\to 
r_{s}/2$, the surface gravity and thus the temperature vanishes. Then in this limit, the expression 
(\ref{eq:ND-nonextremal}) confirms a vanishing number density. Thus taking the extremal limit from 
the nonextremal case, one can see that the Hawking effect ceases to exist. However, there remains a 
persisting debate whether one can compare the extremal black holes to the extremal limit of 
nonextremal ones (see \cite{Liberati:2000sq} and the references therein and the discussions in 
\cite{Vanzo:1995bh,Balbinot:2007kr}). Moreover, starting with an extremal black hole, it is 
observed in the literature \cite{Alvarenga:2003tx, Liberati:2000sq} that the Bogoliubov 
transformation coefficients do not satisfy the consistency condition arising from the commutator 
brackets between the ladder operators related to the ingoing and outgoing field modes. Thus, the 
standard Bogoliubov transformation method remains inconclusive to provide any decisive outcome in 
this matter. On the other hand, using the canonical formulation \cite{Barman:2018ina}, which closely 
mimics the Bogoliubov transformation method, one is able to perceive a vanishing number density of 
Hawking quanta in an extremal Kerr black hole spacetime, considering some simplifications in the 
analysis. Current work aims to understand this vanishing number density without these underlying 
simplifications and precisely point out the deciding factors behind this phenomenon.

\section{Canonical framework}\label{Sec:canonical-framework}

In the original work, \cite{hawking1975} Hawking considered the Bogoliubov transformation between ingoing and outgoing field modes, which are delineated in terms of the respective null coordinates, to perceive the particle creation in a black hole spacetime. However, these null coordinates cannot be used as dynamical variables to describe a true field Hamiltonian, debarring one to get a canonical description of the phenomena. In this regard, in \cite{Barman:2017fzh,Barman:2018ina} the authors considered a set of near-null coordinates, obtained by slightly deforming the null coordinates, and constructed the necessary field Hamiltonians to realize the Hawking effect in the black hole spacetime. This procedure closely mimics the original formulation provided by Hawking. In particular, in \cite{Barman:2018ina} it was shown that using this canonical formulation, one can consistently reproduce the vanishing number density of Hawking quanta in an extremal Kerr black hole spacetime. In this work, we will consider this Hamiltonian formulation and first talk about the near-null coordinates necessary for its understanding.

\subsection{The near-null coordinates}
For an observer near the past null infinity $\scriminus$, say observer $\observerminus$, one defines the near-null coordinates as
\begin{equation}\label{eq:NearNullCoordinatesMinus}
\tau_{-} = t - (1-\epsilon)\rstar ~~;~~ \xi_{-} = -t - (1+\epsilon)\rstar  ~,
\end{equation}
where $\epsilon$ is a very small valued dimensionless real positive parameter such that $\epsilon\gg\epsilon^2$. The near-null coordinates for an observer near the future null infinity $\scriplus$, for observer $\observerplus$, is defined as
\begin{equation}\label{eq:NearNullCoordinatesPlus}
\tau_{+} = t + (1-\epsilon)\rstar ~~;~~ \xi_{+} = -t + (1+\epsilon)\rstar ~.
\end{equation}
We mention that for the past observer $\observerminus$ one has $d\rstar = dr$, as in that case, the 
black hole is not yet formed.

\subsection{Field Hamiltonian and Fourier modes}

From Eq. (\ref{eq:2d-reduced-scalar-action}), one observes that in a Kerr black hole spacetime, the 
reduced scalar field action near the event horizon and the spatial infinities imitate the one from 
$(1+1)$ dimensional flat spacetime, described by the Minkowski metric $ds^2 = -dt^2 + d\rstar^2$.
Then in terms of the near-null coordinates from Eq. (\ref{eq:NearNullCoordinatesMinus}, 
\ref{eq:NearNullCoordinatesPlus}) one can obtain the line-elements corresponding to $\observerminus$ 
and $\observerplus$, related to the conformally transformed flat metric $g^{0}_{\mu\nu}$, as
\begin{equation}\label{eq:NearNullMetricMP}
ds^2_{\pm} = \tfrac{\epsilon}{2} \Big[ - d\tau_{\pm}^2 + d\xi_{\pm}^2 +\tfrac{2}{\epsilon} d\tau_{\pm} d\xi_{\pm} \Big] 
\equiv \tfrac{\epsilon}{2}~ g^{0}_{\mu\nu}dx_{\pm}^{\mu} dx_{\pm}^{\nu}  ~.
\end{equation}
The subscript $+$ and $-$ denote the cases related to observer $\observerplus$ and $\observerminus$, respectively. The reduced scalar field action from
(\ref{eq:2d-reduced-scalar-action}) for both observers can now be expressed as 
\begin{equation}\label{ReducedScalarAction2DFlatMinus}
S_{\varphi} =  \int d\tau_{\pm}  d\xi_{\pm} \left[-\tfrac{1}{2} \sqrt{-g^{0}} 
g^{0\mu\nu} \partial_{\mu}\varphi \partial_{\nu} \varphi \right]  ~.
\end{equation}
Here we have omitted the subscripts from the redefined field $\varphi_{lm}$ for brevity of notation. From Eq. (\ref{eq:NearNullMetricMP}), one sees that corresponding lapse function, shift vector, and the determinant of the spatial metric are respectively given by $N = 1/\epsilon$, $N^1 = 1/\epsilon$, and $q = 1$. Then considering spatial slicing of 
the reduced spacetime with respect to $\tau_{\pm}$, the scalar field Hamiltonian for the two observers can be expressed as
\begin{equation}\label{ScalarHamiltonianOpm}
H_{\varphi}^{\pm} = \int d\xi_{\pm}\; \tfrac{1}{\epsilon}\left[\left\{ 
\tfrac{1}{2} \Pi^2  + \tfrac{1}{2}  (\partial_{\xi_{\pm}}\varphi)^2 \right\} 
+ \Pi~ \partial_{\xi_{\pm}} \varphi \right] ~,
\end{equation}
with $\Pi$ being the conjugate momentum to the field $\varphi$. The momentum $\Pi$ can be obtained from the Hamilton's equation, given by
\begin{equation}\label{FieldMomentumPM}
\Pi(\tau_{\pm},\xi_{\pm}) = \epsilon\;\partial_{\tau_{\pm}}\varphi - 
\partial_{\xi_{\pm}}\varphi ~.
\end{equation}
The field $\varphi$ and the momentum $\Pi$ satisfy the Poisson bracket 
\begin{equation}\label{PoissonBracketPM}
\{\varphi(\tau_{\pm},\xi_{\pm}), \Pi(\tau_{\pm},\xi_{\pm}')\} = 
\delta(\xi_{\pm} - \xi_{\pm}') ~.
\end{equation}
From Eqn. (\ref{ScalarHamiltonianOpm}) one can observe that the Hamiltonian becomes ill-defined at $\epsilon=0$, representing the necessity of near-null coordinates for the realization of the Hawking effect using a Hamiltonian formulation.
%
%
Now, as $\sqrt{q}=1$ and $V_{\pm}=\int d\xi_{\pm}\sqrt{q}$ we consider finite fiducial box during the intermediate steps of computations to avoid dealing with diverging spatial volumes, with
\begin{equation}\label{SpatialVoumePM}
V_{\pm} = \int_{\xi_{\pm}^L}^{\xi_{\pm}^R} d\xi_{\pm}\sqrt{q} = {\xi_{\pm}^R} - 
{\xi_{\pm}^L}  ~,
\end{equation}
and the Fourier transformations of the scalar field for the observers $\observerplus$ and $\observerminus$ are defined as
\begin{eqnarray}
\varphi(\tau_{\pm},\xi_{\pm}) &=& \tfrac{1}{\sqrt{V_{\pm}}}\sum_{k} 
\tilde{\phi}^{\pm}_{k} 
e^{i k \xi_{\pm}}  ~,
\nonumber \\
\Pi(\tau_{\pm},\xi_{\pm}) &=&  \tfrac{1}{\sqrt{V_{\pm}}} \sum_{k} \sqrt{q}~ 
\tilde{\pi}^{\pm}_{k} 
e^{i k \xi_{\pm}} ~,
\label{FourierModesDefinitionPM}
\end{eqnarray}
where the complex-valued mode functions $\tilde{\phi}^{\pm}_{k}$, $\tilde{\pi}^{\pm}_{k}$ depend on $\tau_{\pm}$. Here the Kronecker and Dirac delta are defined as $\int d\xi_{\pm}\sqrt{q} e^{i(k-k')\xi_{\pm}} = V_{\pm} \delta_{k,k'}$ and $\sum_k e^{ik(\xi_{\pm}-\xi_{\pm}')} = V_{\pm} \delta(\xi_{\pm}-\xi_{\pm}')/\sqrt{q}$, and they imply $k\in \{k_s\}$ where $k_s=2\pi s/V_{\pm}$ with $s$ being nonzero integers. With the help of these definitions one can express the field Hamiltonians (\ref{ScalarHamiltonianOpm}) in terms 
of the Fourier modes as $H_{\varphi}^{\pm} = \sum_{k} \tfrac{1}{\epsilon} \left(\mathcal{H}_{k}^{\pm} + \mathcal{D}_{k}^{\pm}\right)$ where
\begin{equation}\label{FourierHamiltonianPM}
\mathcal{H}_{k}^{\pm} = \tfrac{1}{2} \tilde{\pi}^{\pm}_{k}  
\tilde{\pi}^{\pm}_{-k} + \tfrac{1}{2} k^2 \tilde{\phi}^{\pm}_{k} 
\tilde{\phi}^{\pm}_{-k} ~,
\end{equation}
and
\begin{equation}\label{DiffeomorphismGenerator}
\mathcal{D}_{k}^{\pm} =  
-\tfrac{i k}{2} ( \tilde{\pi}^{\pm}_{k} \tilde{\phi}^{\pm}_{-k} -
\tilde{\pi}^{\pm}_{-k} \tilde{\phi}^{\pm}_{k} )  ~,
\end{equation}
respectively denote the Hamiltonian densities and diffeomorphism generators and the corresponding Poisson brackets are now discretized as
\begin{equation}\label{FourierPoissonBracketPlus}
\{\tilde{\phi}^{\pm}_{k}, \tilde{\pi}^{\pm}_{-k'}\} = \delta_{k,k'} ~.
\end{equation}

\subsection{Relation between Fourier modes}

Since we are dealing with scalar field, the fact that $\varphi(\tau_{-},\xi_{-}) = 
\varphi(\tau_{+},\xi_{+})$ makes it possible to express a particular field mode near $\scriplus$ in terms of all the modes near $\scriminus$. Similarly one can show that the field momentum at those regions obey $\Pi(\tau_{+},\xi_{+}) = (\partial \xi_{-}/\partial \xi_{+}) \Pi(\tau_{-},\xi_{-})$ \cite{Barman:2017fzh}. This follows from Eqn.(\ref{FieldMomentumPM}) along with the fact that ingoing and outgoing modes travel keeping the null coordinates $v$ and $u$ constant, respectively. Using these the relations between the Fourier modes and their conjugate momenta follows the relations 
\begin{equation}\label{FieldModesRelation}
\tilde{\phi}^{+}_{\kr} = \sum_{k} \tilde{\phi}^{-}_{k} F_{0}(k,-\kr) ~;~
\tilde{\pi}^{+}_{\kr} =  \sum_{k} \tilde{\pi}^{-}_{k} F_{1}(k,-\kr)  ~,
\end{equation}
where the coefficient functions $F_{n}(k,\kr)$ are given by
\begin{equation}\label{eq:CoeffFunction-General}
F_{n}(k,\kr) = \frac{1}{\sqrt{V_{-} V_{+}}} \int d\xi_{+} 
\left(\tfrac{\partial \xi_{-}}{\partial \xi_{+}}\right)^n
~e^{i k \xi_{-} + i \kr \xi_{+}} ~,
\end{equation}
where $n=0,1$ and these coefficient functions mimic the Bogoliubov coefficients \cite{hawking1975}. Using the general definition of the Dirac delta distribution $\delta(\mu) = \tfrac{1}{2\pi} \int dx~e^{i\mu x}$ and by choosing $\mu=1$, $x=(\pm k\xi_{-} + \kr\xi_{+})$ one gets
\begin{equation}\label{F0F1Relation}
F_{1}(\pm k,\kr) = \mp \left(\tfrac{\kr}{k}\right) F_{0}(\pm k,\kr) ~.
\end{equation}
So the evaluation of only one type of coefficient function, corresponding to $n=0$ or $n=1$, will be sufficient for our purpose.
%
%
Also it is evident that the coefficient functions $F_{n}(\pm k,\pm\kr)$ are complex conjugate to $F_{n}(\mp k,\mp\kr)$. Later on we will use this fact to avoid mathematical complications.

\subsection{Consistency condition and number density of Hawking quanta}

Using Eq. (\ref{F0F1Relation}) and demanding that the two different Poisson brackets $\{\tilde{\phi}^{-}_{k}, \tilde{\pi}^{-}_{-k'}\} = \delta_{k,k'}$ and $\{\tilde{\phi}^{+}_{\kr}, \tilde{\pi}^{+}_{-\kr'}\} = \delta_{\kr,\kr'}$ be simultaneously satisfied, we may express a consistency requirement among the coefficient functions as
\begin{equation}\label{PoissonBracketConsistencyCondition}
\mathbb{S}_{-}(\kr) - \mathbb{S}_{+}(\kr) = 1 ~,
\end{equation}
where $\mathbb{S}_{\pm}(\kr) = \sum_{k>0} (\kr/k) |F_{0}(\pm k,\kr)|^2$. One may find a similarity between this condition and the one from the Bogoliubov transformation method \cite{Alvarenga:2003tx} with the latter one arising from the commutation relation of the ladder operators of the field 
modes.
%
%
Using relations (\ref{FieldModesRelation}) and (\ref{F0F1Relation}) one can express $\mathcal{H}_{\kr}^{+}$ and $\mathcal{D}_{\kr}^{+}$ as 
\begin{equation}\label{ModesHamiltonianRelations0}
\mathcal{H}_{\kr}^{+} = h_{\kr}^1 + \sum_{k>0} \left(\frac{\kr}{k}\right)^2 
\left[\right|F_{0}(-k,\kr)|^2 + |F_{0}(k,\kr)|^2] ~\mathcal{H}_k^{-}  ~,
\end{equation}
\begin{equation}\label{ModesDiffeomorphismRelations0}
\mathcal{D}_{\kr}^{+} = d_{\kr}^1 + \sum_{k>0} \left(\frac{\kr}{k}\right)^2 
\left[\right|F_{0}(-k,\kr)|^2 + |F_{0}(k,\kr)|^2] ~\mathcal{D}_k^{-}  ~,
\end{equation}
where $h_{\kr}^1$ and $d_{\kr}^1$ are linear in $\phi^{-}_{k}$ and 
its conjugate momentum, i.e., the vacuum expectation values of their quantum counterpart vanish.

Now for real-valued scalar field the Fourier modes satisfy $\tilde{\phi}^{*}_{k}=\tilde{\phi}_{-k}$, implying the real and imaginary parts of these Fourier modes not being independent. One may suitably select one of these real or imaginary parts in different domains of the Fourier modes \cite{Hossain:2010eb,Barman:2017fzh} so that the previous reality condition is implemented. It is seen \cite{Hossain:2010eb,Barman:2017fzh} that this makes $\mathcal{D}_k^{-} = 0$ and the Hamiltonian density to represent a simple harmonic oscillator
\begin{equation}\label{RealHamiltonian}
\mathcal{H}_{k}^{\pm} = \frac{1}{2} \pi^{2}_{k}
+ \frac{1}{2} k^2 \phi^{2}_{k} 
~,~~~\{\phi^{2}_{k},\pi^{2}_{k'}\}=\delta_{k,k'}~,
\end{equation}
where $\phi_{k}$ and $\pi_{k}$ are the redefined Fourier field modes which are real-valued.
Now a mode with wave-vector $k$ has frequency $|k|$ and the energy spectrum for each of these oscillator modes is given by $\hat{\mathcal{H}}_{k}^{-}|n_k\rangle =  (\hat{N}_{k}^{-}+\tfrac{1}{2}) |k| |n_k\rangle = (n_k+\tfrac{1}{2}) k |n_k\rangle $ where $\hat{N}_{k}^{-}$ is the corresponding number operator, $|n_k\rangle$ are its eigen-states with integer eigenvalues $n_k \ge 0$. It is also understood that for realizing the Hawking effect one has to evaluate the expectation value of the Hamiltonian density operator 
$\hat{\mathcal{H}}_{\kr}^{+} \equiv (\hat{N}_{\kr}^{+}+\tfrac{1}{2})$ for observer $\observerplus$ in the vacuum state $|0_{-}\rangle = \Pi_{k} |0_k\rangle$ of the observer $\observerminus$. Then using Eq. (\ref{PoissonBracketConsistencyCondition}) with 
(\ref{ModesHamiltonianRelations0}) the expectation value of the number density operator corresponding to the Hawking quanta of frequency $\omegat = \kr$ is obtained as \cite{Barman:2017fzh, Barman:2018ina}
\begin{equation}\label{NumberDensityEVGeneral} 
N_{\omegat} = N_{\kr} \equiv \langle 0_{-}|\hat{N}_{\kr}^{+}|0_{-}\rangle 
=  \mathbb{S}_{+}(\kr)   ~.
\end{equation}

\section{Extremal Kerr black hole}\label{Sec:extremal-kerr-Hawking}

To understand the particle creation in an extremal Kerr black hole spacetime, we have to first determine the relation between the spatial near-null coordinates $\xi_{+}$ and $\xi_{-}$, which requires the expression of the tortoise coordinate from Eq. (\ref{eq:tortoise-extremal}). To derive this relation, we consider a spatial $\tau_{-} = \textup{constant}$ hyper-surface, with a pivotal point $\xi_{-}^0$ on it, corresponding to observer $\observerminus$. Then any spacelike interval can be expressed as
\begin{equation}\label{eq:ximinus-relation}
	(\xi_{-} - \xi_{-}^0)_{|\tau_{-}} = 2(\rstar^0-\rstar)_{|\tau_{-}} =
	2(r^0-r)_{|\tau_{-}} \equiv \Delta ~, 
\end{equation}
where $r^0$ relates to $\xi_{-}^0$. On the other hand, corresponding to observer $\observerplus$ a spacelike  interval on a $\tau_{+} = \textup{constant}$ hyper-surface can be expressed using Eqn. (\ref{eq:tortoise-extremal}), as
\begin{equation}\label{eq:xiplus-relation}
	(\xi_{+} - \xi_{+}^0)_{|\tau_{+}} = \Delta 
	+ 2r_{s} \ln \left(1 + \tfrac{\Delta}{\Delta_0}\right) - 
	\tfrac{2r_{s}^2}{\Delta+\Delta_0} + \tfrac{2r_{s}^2}{\Delta_0}~,
\end{equation}
where using geometric optics approximation one can identify the interval $2(r - r^0)_{|\tau_{+}}$ as $\Delta$ and we have $\Delta_0 \equiv 2(r^0 - r_{s}/2)_{|\tau_{+}}$. Furthermore, choosing the pivotal value $\xi_{-}^0 = \Delta_0$ and $\xi_{+}^0 = \xi_{-}^0 + 2r_{s} \ln (\xi_{-}^0/\sqrt{2} r_{s})-2r_{s}^2/\xi_{-}^0$, one can express the above relation as
\begin{equation}\label{eq:near-null-relation-extremal}
	\xi_{+} = \xi_{-} + 2r_{s} \ln \left(\frac{\xi_{-}}{\sqrt{2} 
		r_{s}}\right)-\frac{2 r_{s}^2}{\xi_{-}} ~.
\end{equation}
The presence of the inverse term makes the relation qualitatively different from nonextremal case \cite{Barman:2018ina}. Furthermore, to keep track of contribution of the individual term in the right hand side of Eq. (\ref{eq:near-null-relation-extremal}) we introduce the parameters $\alpha_{j}$, with $j=1,2,3$, and express the relation as
\begin{equation}\label{eq:near-null-relation-extremal-f}
\xi_{+} = \alpha_{1}~ \xi_{-} + \alpha_{2}~ 2r_{s} \ln \left(\frac{\xi_{-}}{\sqrt{2} r_{s}}\right)-\alpha_{3}~ \frac{2 r_{s}^2}{\xi_{-}} ~,
\end{equation}
where $\alpha_{j}$ can only take values $0$ or $1$. In our subsequent analysis we shall be using this relation to evaluate the consistency condition and the number density of Hawking quanta in an extremal Kerr black hole spacetime. Later we shall also consider only the inverse term in the relation (\ref{eq:near-null-relation-extremal-f}) to understand the consequences.

\subsection{Understanding the Hawking effect using the complete relation among the spatial near-null coordinates}
Here we shall be evaluating the coefficient functions of Eq. (\ref{eq:CoeffFunction-General}) with the complete relation between the spatial near-null coordinates $\xi_{+}$ and $\xi_{-}$ from Eq. (\ref{eq:near-null-relation-extremal-f}). Then we shall give a detailed understanding on the consistency condition from Eq. (\ref{PoissonBracketConsistencyCondition}) and subsequently the number density (\ref{NumberDensityEVGeneral}) corresponding to the Hawking effect. It should be noted that in the general relation (\ref{eq:near-null-relation-extremal-f}) by keeping only a certain $\alpha_{j}$ non vanishing one can ascertain the role of individual terms contributing in the satisfaction of the consistency condition and also in the vanishing number density of the Hawking quanta. In our following study we shall also discuss few approximations on the relation (\ref{eq:near-null-relation-extremal-f}) and their outcomes.

\subsubsection{Evaluation of the coefficient functions}

As already discussed one can either estimate $F_{0}(\pm k,\kr)$ or $F_{1}(\pm k,\kr)$ and get the expression of the other coefficient function from (\ref{F0F1Relation}). In particular, here for the lucidity of calculation we estimate only $F_{1}(\pm k,\kr)$. $F_{0}(\pm k,\kr)$ will follow from relation (\ref{F0F1Relation}). Putting the expression of $\xi_{+}$ from Eq. (\ref{eq:near-null-relation-extremal-f}) in Eq. (\ref{eq:CoeffFunction-General}) with $n=1$ we get
\begin{eqnarray}\label{F1withCompleteRelation}
F_{1}(\pm k,\kr) = \frac{1}{\sqrt{V_{-} V_{+}}} \int_{\xi^L_{-}}^{\xi^R_{-}} d\xi_{-} \Big(\frac{\xi_{-}}{\sqrt{2}r_{s}}\Big)^{i\alpha_{2}2\kappa r_{s}} ~~~~\nonumber \\
\times ~e^{i(\pm k+\alpha_{1}\kappa)\xi_{-}-\frac{i\alpha_{3}2\kappa r_{s}^2}{\xi_{-}}} ~.
\end{eqnarray}
We mention that this integrand is oscillatory in nature and for the limit 
$\xi^L_{-}\to 0$, $\xi^R_{-}\to\infty$ it is essentially divergent. Therefore, in this case we shall be introducing regulators with parameter $``\delta"$ to properly evaluate the integrals
\begin{equation}\label{F1IntegralwithRegulator}
F_{1}^\delta(\pm k,\kr) = \frac{b}{\sqrt{V_{-} V_{+}}} \int_{x^{L}}^{x^{R}} dx 
~x^{\sqrt{2}\alpha_{2}b_{0}} \exp\Big(-b_{\pm}x-\frac{\alpha_{3}b_0}{x}\Big)~,
\end{equation}
where $b=\sqrt{2}\rs$, $b_{\pm} = b[\delta|\pm k+\alpha_{1}\kappa|-i(\pm k+\alpha_{1}\kappa)]$ and $b_0 = b[\delta|\kappa|+i\kappa]$. Here we have represented the original integral in terms of the dimensionless variable $x=\xi_{-}/b$.
%
%
It should be noted that in the limit $\delta\rightarrow0$ Eq. (\ref{F1IntegralwithRegulator}) goes back to Eq. (\ref{F1withCompleteRelation}). 

One can notice from Eq. (\ref{F1withCompleteRelation}) that there is a possibility of $b_{-}=0$ when $\alpha_{1}\kappa=k$. In that scenario the characteristic form of the integral will be different. So we will explore this particular situation separately. Also, we remind that $k,\kappa>0$. So this particular situation will not arise if $\alpha_{1} = 0$. It is worth mentioning that the fact $F_{n}(\pm k,\kappa)$ is complex conjugate to $F_{n}(\mp k,-\kappa)$ does not get violated due to the inclusion of the regularization parameter $\delta$. 

\paragraph{\underline{Evaluation of $F_{1}(-\alpha_{1}\kappa,\kappa)$}}:
In this particular situation $\alpha_{1}\kappa=k$ and $b_{-}=0$. Then the integral form Eq. (\ref{F1withCompleteRelation}) becomes
\begin{equation}\label{eq:F1-complete-a-1}
F_{1}^{\delta}(-\alpha_{1}\kappa,\kappa) = \frac{1}{\sqrt{V_{-} V_{+}}} \int_{\xi_{-}^L}^{\xi_{-}^R} d\xi_{-} \Big(\frac{\xi_{-}}{\sqrt{2}r_{s}}\Big)^{i\alpha_{2}2\kappa r_{s}} e^{-\frac{i\alpha_{3}2\kappa r_{s}^2}{\xi_{-}}} ~.
\end{equation}
In terms of the dimensionless variable along with the $\delta$ regulator, i.e., the expression from Eq. (\ref{F1IntegralwithRegulator}) in this scenario becomes
\begin{equation}\label{eq:F1-complete-a-2}
F_{1}^{\delta}(-\alpha_{1}\kappa,\kappa) = \frac{b}{\sqrt{V_{-} V_{+}}} \int_{x^L}^{x^R} dx~x^{\sqrt{2}\alpha_{2}b_{0}} e^{-\frac{\alpha_{3}b_{0}}{x}}~.
\end{equation}
The explicit form of the above integral can be given in terms of 
\textit{incomplete gamma} functions as
\begin{eqnarray}\label{eq:F1-complete-a-3}
F_{1}^{\delta}(-\alpha_{1}\kappa,\kappa) &=& 
\frac{b(\alpha_{3}b_{0})^{\sqrt{2}\alpha_{2}b_{0}+1}}{\sqrt{V_{-} 
V_{+}}}\Big[\Gamma\Big(-1-\sqrt{2}\alpha_{2}b_{0},\frac{\alpha_{3}b_{0}}{x^{R}
}\Big) \nonumber\\ 
~&&~~~~~~ -\Gamma\Big(-1-\sqrt{2}\alpha_{2}b_{0},\frac{\alpha_{3}b_{0}}{x^{L}}
\Big)\Big ]~. 
\end{eqnarray}
The second \textit{gamma} function within the square bracket will vanish as 
$x^{L}\rightarrow 0$. Following the properties of \textit{incomplete gamma 
functions} \cite{NIST:DLMF} in the limit  $x^{R}\rightarrow\infty$, we get the above expression to become
\begin{eqnarray}\label{eq:F1square-complete-a}
|F_{1}^{\delta}(-\alpha_{1}\kappa,\kappa)|^2 \approx \frac{\gamma}{1+2\alpha^2_{2} b^2\kr^2}~,
\end{eqnarray}
where $\gamma \equiv V_{-}/V_{+}$. In deriving the above result we have used the 
fact that in the limit $x^{L}\rightarrow 0~,~x^{R}\rightarrow\infty$, $V_{-}\sim 
x^{R}$.

\paragraph{\underline{Evaluation of $F_{1}(\pm k,\kappa)$, for $k\neq\alpha_{1}\kappa$}}:
When $k\neq\alpha_{1}\kappa$ one has $b_{-}\neq0$, and the expression of the 
integral (\ref{F1IntegralwithRegulator}) can be obtained in terms of 
\textit{modified Bessel} functions of second kind $\mathcal{K}(\mu,z)$ as
\begin{eqnarray}\label{eq:F1-complete-b-1}
F_{1}^\delta(\pm k,\kappa) &=& \frac{2b}{\sqrt{V_{-} 
V_{+}}}\Big(\frac{\alpha_{3}b_{0}}{b_{\pm}}\Big)^{\frac{1+\sqrt{2}\alpha_{2}b_{0}}{2}} \nonumber\\ 
&\times&\mathcal{K}(-1-\sqrt{2}\alpha_{2}b_{0},2\sqrt{\alpha_{3}b_{\pm}b_{0}} 
)~.
\end{eqnarray}
Here we have considered the limits of the integral 
(\ref{F1IntegralwithRegulator}) from $0$ to $\infty$ instead of $x^{L}$ to 
$x^{R}$, which can be done by adding two boundary terms which tends to zero as 
$x^{L}\rightarrow0$ and $x^{R}\rightarrow\infty$ \cite{Barman:2018ina}. We mention that 
these modified Bessel functions have the following approximate expressions \cite{NIST:DLMF} for 
different limits of their arguments
\begin{eqnarray}\label{ModifiedBesselFunctionLimitingBehavior}
\mathcal{K}(\mu,z) &\sim& \frac{1}{2}\Big(\frac{2}{z}\Big)^{\mu}\Gamma(\mu)~; 
~~~\text{as}~~~ z\rightarrow 0~,\nonumber\\ \mathcal{K}(\mu,z) &\sim& 
\sqrt{\frac{\pi}{2z}} e^{-z}~; ~~~\text{as}~~~ z\rightarrow \infty~. 
\end{eqnarray}
Here $\mu$ and $z$ can be both real as well as complex valued. Another relation 
satisfied by these modified Bessel functions is $\mathcal{K}(\mu,z) = 
\mathcal{K}(-\mu,z)$, which will also be relevant in our study.

\subsubsection{Consistency condition}
So far we have determined the coefficient functions. Now we proceed to check for 
the consistency condition and the number density of the Hawking quanta. In this 
regard, first we need to estimate the LHS of 
Eq. (\ref{PoissonBracketConsistencyCondition}) which reads 
\begin{eqnarray}\label{eq:Consistency-Cond-complete-1}
	\mathbb{S}_{-}^{\delta}(\kappa) - \mathbb{S}_{+}^{\delta}(\kappa) &=& |F_{1}^{\delta}(-\alpha_{1}\kappa,\kappa)|^2 \nonumber\\
	~&+&  \sum_{\substack{k>0\\k\neq\alpha_{1}\kappa}} \Big(\frac{k}{\kappa}\Big) |F_{1}^{\delta}(- k,\kappa)|^2  \nonumber\\ 
	~&-& \sum_{k>0} \Big(\frac{k}{\kappa}\Big) |F_{1}^{\delta}(k,\kappa)|^2 ~.
\end{eqnarray}
An explicit calculation using the functional form of the coefficient function 
$F_{1}^{\delta}$, as derived in Eq. (\ref{eq:F1-complete-b-1}), leads to the 
following
\begin{eqnarray}\label{eq:F1square-complete}
|F_{1}^{\delta}(k,\kappa)|^2 &\approx& \tfrac{e^{-\pi\sqrt{2}b\kr\alpha_{2}}}{V_{-} V_{+}}  \frac{1}{(\alpha_{1}\kappa+k)^2} |z_{+}\mathcal{K}(\bar{\mu},z_{+})|^2 \nonumber\\
|F_{1}^{\delta}(-k,\kappa)|^2 &\approx& \tfrac{e^{-\pi\sqrt{2}b\kr\alpha_{2}}}{V_{-} V_{+}}  \tfrac{1}{(\alpha_{1}\kappa-k)^2} |z_{-}\mathcal{K}(\bar{\mu},z_{-})|^2 ~;\alpha_{1}\kappa>k \nonumber\\
 &=& \frac{1}{V_{-} V_{+}}  \frac{1}{(k-\alpha_{1}\kappa)^2} |z_{-}\mathcal{K}(\bar{\mu},z_{-})|^2 ~;\alpha_{1}\kappa<k~,\nonumber\\
\end{eqnarray}
where $\bar{\mu}=-1-\sqrt{2}\alpha_{2}b_{0}$. Also, $z_{-}$ for $\kappa > k$ is not same as $z_{-}$ for $\kappa < k$ and $z_{\pm}$ are given by

\begin{eqnarray}\label{eq:Zpm-complete}
z_{\pm} &=& 2\sqrt{\alpha_{3}b_{0}b_{\pm}}\nonumber\\
~&=& 2\sqrt{\frac{\alpha_{3}|b_{0}|^2(i+\delta)\big[\delta|\alpha_{1}\mstar\pm 
s|-i(\alpha_{1}\mstar\pm s)\big]}{\mstar}}~. \nonumber\\
\end{eqnarray}
Here we have used $k=2\pi s/V_{-}$ and $\mstar = V_{-}\kappa/(2\pi)$. The first term in the RHS of Eq. (\ref{eq:Consistency-Cond-complete-1}) has already been calculated previously in Eq. (\ref{eq:F1square-complete-a}). Now we shall concentrate in calculating the other two summations. Using Eq. (\ref{eq:F1square-complete}) and (\ref{eq:Zpm-complete}) the last two terms of the RHS of (\ref{eq:Consistency-Cond-complete-1}) becomes
\begin{widetext}
\begin{eqnarray}\label{eq:Consistency-Cond-complete-rel-2}
\sum_{\substack{k>0\\k\neq\alpha_{1}\kappa}} \Big(\frac{k}{\kappa}\Big) |F_{1}^{\delta}(- k,\kappa)|^2  &-& \sum_{k>0} \Big(\frac{k}{\kappa}\Big) |F_{1}^{\delta}(k,\kappa)|^2 
= \frac{\gamma}{4\pi^2 \mstar}\Big[ e^{-\pi\sqrt{2}b\kr\alpha_{2}} \sum_{\substack{s=1\\s<\alpha_{1}\mstar}}^{\alpha_{1}\mstar-1}\frac{s}{(\alpha_{1}\mstar-s)^2}|z_{-}\mathcal{K}(\bar{\mu},z_{-})|^2 \nonumber\\
&+&  \sum_{\substack{s=\alpha_{1}\mstar+1\\s>\alpha_{1}\mstar}}^{\infty}\frac{s}{(s-\alpha_{1}\mstar)^2}|z_{-}\mathcal{K}(\bar{\mu},z_{-})|^2 - e^{-\pi\sqrt{2}b\kr\alpha_{2}} \sum_{s=1}^{\infty}\frac{s}{(\alpha_{1}\mstar+s)^2}|z_{+}\mathcal{K}(\bar{\mu},z_{+})|^2\Big]~.
\end{eqnarray}
It is worth mentioning that $z_{-}$ in the first and second summation is not exactly the same but it has to be calculated for $s<\alpha_{1}\mstar$ and $s>\alpha_{1}\mstar$, respectively. Performing a change in variables as $s-\alpha_{1}\mstar = p,~\alpha_{1}\mstar-s = q$ and $s+\alpha_{1}\mstar = r$, the above summation can be recast in the following way
\begin{eqnarray}\label{eq:Consistency-Cond-complete-rel-3}
\frac{\gamma}{4\pi^2 \mstar}\Big[ e^{-\pi\sqrt{2}b\kr\alpha_{2}} \sum_{q=1}^{\alpha_{1}\mstar-1} \frac{\alpha_{1}\mstar-q}{q^2}|\tilde{z}(q) \mathcal{K}(\bar{\mu},|\tilde{z}(q)|)|^2 &+& \sum_{p=1}^{\infty} \frac{\alpha_{1}\mstar+p}{p^2}|\tilde{z}(p) \mathcal{K}(\bar{\mu},\tilde{z}(p))|^2 \nonumber\\
&-& e^{-\pi\sqrt{2}b\kr\alpha_{2}} \sum_{r=\alpha_{1}\mstar+1}^{\infty} \frac{r-\alpha_{1}\mstar}{r^2}|\tilde{z}(r) \mathcal{K}(\bar{\mu},|\tilde{z}(r)|)|^2\Big]~.
\end{eqnarray}
Here $z_{-}(s)$ has been changed to $|\tilde{z}(q)|=\sqrt{4\alpha_{3}|b_{0}|^2 q/\mstar}$ for $\alpha_{1}\mstar>s$. Similarly for $\alpha_{1}\mstar<s$, $z_{-}(s)$ takes the form  $\tilde{z}(p)=\sqrt{4\alpha_{3}|b_{0}|^2 p/\mstar}(i+\delta)$. Also, $z_{+}(s)$ goes to $|\tilde{z}(r)|=\sqrt{4\alpha_{3}|b_{0}|^2 r/\mstar}$. Rearranging the above summations leads to
\begin{eqnarray}\label{eq:Consistency-Cond-complete-rel-4}
\frac{\gamma}{4\pi^2} \sum_{s=1}^{\infty} \frac{\alpha_{1}}{s^2}\Big[ \big|\tilde{z}(s)\mathcal{K}(\bar{\mu},\tilde{z}(s))\big|^2 + e^{-\pi\sqrt{2}b\kr\alpha_{2}} \big|\tilde{z}(s)\mathcal{K}(\bar{\mu},|\tilde{z}(s)|)\big|^2 \Big] &+& \frac{\gamma}{4\pi^2 \mstar} \sum_{s=1}^{\infty} \frac{1}{s}\Big[ |\tilde{z}(s)\mathcal{K}(\bar{\mu},\tilde{z}(s))|^2 \nonumber\\
&-& e^{-\pi\sqrt{2}b\kr\alpha_{2}} |\tilde{z}(s)\mathcal{K}(\bar{\mu},|\tilde{z}(s)|)|^2 \Big] ~,
\end{eqnarray}
\end{widetext}
where $\tilde{z}(s) = \sqrt{4\alpha_{3}|b_{0}|^2 s/\mstar}(i+\delta)$, and we also note that 
$\pi\sqrt{2}b\kr\alpha_{2}=2\pi\rs\kr\alpha_{2}$. So far we have not used any asymptotic expressions 
(\ref{ModifiedBesselFunctionLimitingBehavior}) of the Bessel function. From the above expression it 
is clear that in the infinite volume limit \textit{i.e.,} $V_{-}\rightarrow \infty$ which implies 
$\mstar\rightarrow\infty$, the second summation vanishes. Now using the approximations for modified 
Bessel functions, as mentioned in Eq. (\ref{ModifiedBesselFunctionLimitingBehavior}), when 
$\tilde{z}\rightarrow\ 0$, and $\mathcal{K}(-\bar{\mu},z) = \mathcal{K}(\bar{\mu},z)$ we get
\begin{eqnarray}\label{eq:zK^2-exps-1}
|\tilde{z}(s)\mathcal{K}(\bar{\mu},\tilde{z}(s))|^2 &\approx& \frac{2\pi\rs\kappa\alpha_{2} ~e^{2\pi\rs\kappa\alpha_{2}}}{\sinh(2\pi\rs\kappa\alpha_{2})} \nonumber\\
|\tilde{z}(s)\mathcal{K}(\bar{\mu},|\tilde{z}(s)|)|^2 &\approx& \frac{2\pi\rs\kappa\alpha_{2}}{\sinh(2\pi\rs\kappa\alpha_{2})}~.
\end{eqnarray}

Then from these expressions also one can observe that the second term in Eq. (\ref{eq:Consistency-Cond-complete-rel-4}) vanishes when $\tilde{z}\to 0$ and one takes $\mstar\to\infty$. On the other hand, when $\tilde{z}\to\ \infty$ one can get the limiting expressions as
\begin{eqnarray}\label{eq:zK^2-exps-2}
|\tilde{z}(s)\mathcal{K}(\bar{\mu},\tilde{z}(s))|^2 &\approx& \sqrt{\frac{\alpha_{3}\pi^2 |b_{0}|^2 s}{\mstar}} e^{-2\delta\sqrt{\frac{4\alpha_{3} |b_{0}|^2 s}{\mstar}}} \nonumber\\
|\tilde{z}(s)\mathcal{K}(\bar{\mu},|\tilde{z}(s)|)|^2 &\approx& \sqrt{\frac{\alpha_{3}\pi^2 |b_{0}|^2 s}{\mstar}} e^{-2\sqrt{\frac{4\alpha_{3} |b_{0}|^2 s}{\mstar}}}~.
\end{eqnarray}
These expressions also confirm that the second term in Eq. (\ref{eq:Consistency-Cond-complete-rel-4}) vanishes when $\tilde{z}\to \infty$ as one takes $\mstar\to\infty$. Here in the first equation one can easily see that in the case $\delta=0$ the right hand side term diverges out as $s\rightarrow\infty$. Finally, the LHS of the consistency relation Eq. (\ref{PoissonBracketConsistencyCondition}) reads

\begin{widetext}

\begin{eqnarray}\label{eq:Consistency-Cond-complete-rel-5}
	\mathbb{S}_{-}^{\delta}(\kappa) - \mathbb{S}_{+}^{\delta}(\kappa) &=& \frac{\gamma}{1+2\alpha^2_{2} b^2\kr^2} +  \frac{\gamma}{4\pi^2} \sum_{s=1}^{\infty} \frac{\alpha_{1}}{s^2}\Big[ \big|\tilde{z}(s)\mathcal{K}(\bar{\mu},\tilde{z}(s))\big|^2 + e^{-2\pi\rs\kr\alpha_{2}} \big|\tilde{z}(s)\mathcal{K}(\bar{\mu},|\tilde{z}(s)|)\big|^2 \Big] \nonumber\\
	&\approx& \frac{\gamma}{1+2\alpha^2_{2} b^2\kr^2} + \frac{\alpha_{1}\gamma}{4\pi^2}~ y \coth(y) \int_{1}^{\lambda_{1}\mstar}\frac{ds}{s^2} + \frac{\alpha_{1}\gamma}{4\pi^2}\pi a \int_{\lambda_2 \mstar}^{\infty} \frac{ds}{s^2}\sqrt{s}\Big[e^{-4\delta a\sqrt{s}}+e^{-4 a\sqrt{s}}\Big]~,
\end{eqnarray}
\end{widetext}
where $y=4\pi\rs\kr\alpha_{2}$ and $a=\sqrt{\frac{\alpha_{3}|b_{0}|^2}{\mstar}}$. In order to arrive at the above expression we have considered $\tilde{z} \ll 1$ when $s\in [1,\lambda_{1}\mstar]$ and $\tilde{z} \gg 1$ when $s\in [\lambda_{2}\mstar,\infty)$. After carrying out these integrals and then taking the infinite volume limit, $\mstar\rightarrow\infty$, the consistency relation becomes
\begin{equation}\label{eq:Consist-cond-final}
    \gamma\bigg[\frac{1}{1+4\alpha^2_{2} \rs^2\kr^2} + \frac{\alpha_{1}}{4\pi^2}~ y \coth(y)\bigg] =1~.
\end{equation}
One can notice that this consistency condition does not contain the integral regulator $\delta$, but it depends on the volume regulators $V_{+}$ and $V_{-}$ through the expression of $\gamma$. Then this consistency condition basically says that these two volume regulators corresponding two different observers, namely observer $\observerplus$ and $\observerminus$, are not independent of each other and are related among themselves for a proper consistency of the estimations. In particular, it specifically ascertains that for a fixed frequency of the outgoing field mode $\kr$, the volume $V_{-}$ must be proportional to $V_{+}$, i.e., $V_{-}\propto V_{+}$. From Eq. (\ref{eq:Consist-cond-final}) we also observe that if one takes the case $\alpha_{1}=1$ with the limit $\alpha_{2}\to 0$, i.e., in the case $\lim_{y\to0}[y~\coth(y)]=1$, the consistency condition exactly reduces to $\gamma=1+1/12=13/12$ stating the same requirement for consistency. This latter case signifies making the contribution of the logarithmic term to vanish in the relation between the spatial near-null coordinates of Eq.  (\ref{eq:near-null-relation-extremal-f}), which was obtained in \cite{Barman:2018ina}. Therefore, from our present calculations one can get back the results of \cite{Barman:2018ina} in a straightforward manner.

\subsubsection{Number density of Hawking quanta}

As already observed in Eq. (\ref{NumberDensityEVGeneral}) the expectation value of the number density operator corresponding to the Hawking quanta is give by $N_{\omegat} = \mathbb{S}_{+}(\kr)$, which in the present scenario can be expressed in the form
\begin{eqnarray}\label{eq:number-den-1}
N_{\omegat} &=& \sum_{k>0} \left(\frac{k}{\kr}\right) |F_{1}^{\delta}(k,\kr)|^2\nonumber\\
~&=& \frac{\gamma~e^{-\pi\sqrt{2}b\kr\alpha_{2}}}{4\pi^2\mstar} \sum_{s=1}^{\infty} \frac{s}{(s+\alpha_{1}\mstar)^2} \nonumber\\
~&&~~~~~~~ |z_{+}(s)~\mathcal{K}(-1-\sqrt{2}\alpha_{2}b_{0},z_{+}(s))|^2\nonumber\\
~&=& \frac{\gamma~e^{-\pi\sqrt{2}b\kr\alpha_{2}}}{4\pi^2\mstar} \sum_{s=1+\alpha_{1}\mstar}^{\infty} \frac{s-\alpha_{1}\mstar}{s^2}\nonumber\\
~&& ~~~~~~~~~|\tilde{z}(s)~\mathcal{K}(-1-\sqrt{2}\alpha_{2}b_{0},|\tilde{z}(s)|)|^2~,
\end{eqnarray}
where, $|\tilde{z}(s)|=\sqrt{4\alpha_{3}|b_{0}|^2 s/\mstar}$. This expression of number density from Eq. (\ref{eq:number-den-1}) can be expressed in terms of the sum of two quantities. First, a sum over a quantity with a factor of $1/s$ in it. Second, the sum over a quantity with a multiplicative factor of $1/s^2$ in it. It should be mentioned that there is a term $1/\mstar$ multiplied with the first term, which makes the term vanish in the limit of $\mstar\to\infty$, i.e., in the infinite volume limit. On the other hand, when $\alpha_{1}=0$ the second term vanishes. Let us evaluate the second sum when $\alpha_{1}\neq0$. One can express this sum as
\begin{eqnarray}
~&& \frac{\alpha_{1}\gamma~e^{-\pi\sqrt{2}b\kr\alpha_{2}}}{4\pi^2} \sum_{s=1+\alpha_{1}\mstar}^{\infty} \frac{1}{s^2} |\tilde{z}~\mathcal{K}(-1-\alpha_{2}b_{1},|\tilde{z}|)|^2 \nonumber\\
~&=& \frac{\alpha_{1}\gamma~e^{-\pi\sqrt{2}b\kr\alpha_{2}}}{4\pi^2} \bigg[\sum_{s=1+\alpha_{1}\mstar}^{\lambda\mstar-1} \frac{1}{s^2} |\tilde{z}~\mathcal{K}(-1-\alpha_{2}b_{1},|\tilde{z}|)|^2 \nonumber\\
~&&~~~~~~~~~\sum_{s=\lambda\mstar}^{\infty} \frac{1}{s^2} |\tilde{z}~\mathcal{K}(-1-\alpha_{2}b_{1},|\tilde{z}|)|^2\bigg] \nonumber\\
~&\approx& \frac{\alpha_{1}\gamma~e^{-\pi\sqrt{2}b\kr\alpha_{2}}}{4\pi^2} \bigg[d_{1} \int_{1+\alpha_{1}\mstar}^{\lambda\mstar-1} \frac{ds}{s^2} \nonumber\\
~&& ~~~~~~~~+~ \int_{\lambda\mstar}^{\infty} \frac{ds}{s^2} \left(\frac{\pi|\tilde{z}|}{2}\right)e^{-2|\tilde{z}|}\bigg]~,
\end{eqnarray}
which after carrying out the integration and in the limit of $\mstar\to\infty$ becomes $0$. This implies that $N_{\omegat} = \lim_{\mstar\to\infty}\mathbb{S}_{+}(\kr)=0$.
%
%
Therefore, it can be shown that the number density of the Hawking quanta in an extremal Kerr black hole spacetime vanishes  is the number density of the Hawking quanta in is zero.

We mention that in Eq. (\ref{eq:near-null-relation-extremal-f}) if one puts $\alpha_{2}=1$, and  $\alpha_{1}=0=\alpha_{3}$ then the relation becomes same as the ones considered in \cite{Barman:2017fzh} for Schwarzschild and in \cite{Barman:2018ina} for nonextremal Kerr black holes. This relation is bound to give a Planckian distribution of particles. Therefore, only the logarithmic term in the relation (\ref{eq:near-null-relation-extremal-f}) is incapable to provide vanishing number density of Hawking quanta in an extremal Kerr black hole spacetime. In the subsequent study we shall specifically choose the inverse term in the relation (\ref{eq:near-null-relation-extremal-f}) to understand the consequences.

\subsection{Inverse relation approximation $\xi_{+}\approx -\frac{2r_{s}^{2}}{\xi_{-}}$}

In the preceding studies we have performed the calculations with a complete relation 
(\ref{eq:near-null-relation-extremal-f}) between $\xi_{+}$ and $\xi_{-}$ and we have seen that the 
expectation value of the number density operator corresponding to the Hawking effect comes out to be 
zero in a consistent manner. However, this analysis along with the one in \cite{Barman:2018ina} do 
not explicitly present the situation when the relation between the spatial near-null coordinates 
contains only the inverse term. On the other hand, from our current calculation of previous 
subsections we have already observed that this inverse term in the full relation dominates over the 
other two terms in providing the vanishing number density. Moreover, in literature also it is widely 
believed that similar inverse terms are responsible for vanishing number density of Hawking quanta 
for extremal black holes. But there are certain ambiguities regarding the semi-classical approach 
\cite{Alvarenga:2003tx}. This motivates us to perform a precise investigation keeping only the 
inverse term in the relation (\ref{eq:near-null-relation-extremal-f}), \textit{i.e.,} by setting 
$\alpha_{1}=\alpha_{2}=0$ and $\alpha_{3}=1$. 
It is to be noted that using the entire relation of Eq. (\ref{eq:near-null-relation-extremal-f}) with $\alpha_{1}$, $\alpha_{2}$, and $\alpha_{3}$ all set to $1$, as one takes the limits $\xi_{-}\to0$ and $\xi_{-}\to\infty$ the spatial near-null coordinate $\xi_{+}$ respectively gives $-\infty$ and $\infty$. This remains true even if one sets $\alpha_{2}=0$. On the other hand, with only the inverse relation between the near null coordinates, i.e., when only $\alpha_{3}=1$, one gets the corresponding limits of $\xi_{+}$ as $-\infty$ and $0$. It signifies the difference of this particular case. Here one may expect that the relation (\ref{F0F1Relation}) will not hold true as $\xi_{+}$ does not have a full range $(-\infty,\infty)$ for the considered range of $\xi_{-}$. Therefore, we shall proceed with caution in this case and evaluate both $F_{0}(\pm k,\kr)$ and $F_{1}(\pm k,\kr)$ for our study.

Using the general expression from Eq. (\ref{eq:CoeffFunction-General}) the explicit forms of the coefficient functions using this inverse relation reads
\begin{eqnarray}\label{eq:F01-inverse-1}
	F_{n}(\pm k,\kr) = \frac{1}{\sqrt{V_{-} V_{+}}} \int_{\xi^{L}_{-}}^{\xi^{R}_{-}} d\xi_{-} ~\Big(\frac{2r_{s}^{2}}{\xi_{-}^{2}}\Big)^{1-n}~ e^{\pm i k\xi_{-} - \frac{i2\kappa r_{s}^{2}}{\xi_{-}}}~.
\end{eqnarray}
Here $n$ can only take values $0$ or $1$. In terms of the dimensionless variable $x=\xi_{-}/b=\xi_{-}/\sqrt{2}r_{s}$, this integral can be represented as
\begin{eqnarray}\label{eq:F01-inverse-2}
	F_{n}^{\delta}(\pm k,\kr) = \frac{b}{\sqrt{V_{-} V_{+}}} \int_{x^{L}}^{x^{R}} dx~\bigg(\frac{1}{x^2}\bigg)^{1-n}~e^{-b_{\pm}x - \frac{b_{0}}{x}}~. 
\end{eqnarray}
The above integral of Eq. (\ref{eq:F01-inverse-1}) is formally divergent for both of the $n$ values. We remove that divergence by introducing a $\delta$ regulator in  $b_{\pm}=b[\delta|\pm k|-i(\pm k)]$ and $b_{0}= b[\delta|\kappa|+i\kappa]$. Performing the above integration one gets
\begin{eqnarray}\label{eq:F01-inverse-3}
	F_{0}^{\delta}(\pm k,\kr) &=&  \frac{2b}{\sqrt{V_{-} V_{+}}} \sqrt{\frac{b_{\pm}}{b_{0}}} \mathcal{K}(1, 2\sqrt{b_{\pm}b_{0}})  \nonumber        \\
	F_{1}^{\delta}(\pm k,\kr) &=&  \frac{2b}{\sqrt{V_{-} V_{+}}} \sqrt{\frac{b_{0}}{b_{\pm}}} \mathcal{K}(1, 2\sqrt{b_{\pm}b_{0}})~.
\end{eqnarray}
However, from the above expressions one can easily find a relation between the coefficient functions and this reads
\begin{equation}\label{eq:F0-F1-rel-inverse}
	F_{0}^{\delta}(\pm k,\kr) = \frac{k}{\kappa} \Big(\frac{\mp i +\delta}{i+\delta}\Big) F_{1}^{\delta}(\pm k,\kr) ~.
\end{equation}
In the limit $\delta\rightarrow 0$ the above relation takes the form of Eq. (\ref{F0F1Relation}). 

\subsubsection{Consistency condition}
Here using the expressions of these coefficient functions we shall look into the consistency condition of Eq. (\ref{PoissonBracketConsistencyCondition}). In particular, the LHS of Eq. (\ref{PoissonBracketConsistencyCondition}) can be expressed as 
\begin{eqnarray}\label{eq:Consistency-Cond-inverse-1}
    \mathbb{S}_{-}(\kr) - \mathbb{S}_{+}(\kr) = \sum_{k>0} \Big(\frac{k}{\kappa}\Big) \Big[|F_{1}^{\delta}(- k,\kappa)|^2  
	-  |F_{1}^{\delta}(k,\kappa)|^2\Big] ~.
\end{eqnarray}
We mention that unlike the previous situation, here we shall not get any special term for $k=-\kappa$. Now using Eq. (\ref{eq:F01-inverse-3}), one can easily show that
\begin{equation}\label{InvRelCoeff}
  \Big(\frac{k}{\kappa}\Big)|F_{1}^{\delta}(\pm k,\kappa)|^2  = \frac{\gamma}{4\pi^2 \mstar}\frac{1}{s} |z_{\pm}\mathcal{K}(1, \bar{z}_{\pm})|^2~,
\end{equation}
where $\bar{z}_{-}=2\sqrt{|b_{0}|^2 s/\mstar}(\delta+i)$, $\bar{z}_{+}=|\bar{z}_{-}|=2\sqrt{|b_{0}|^2 s/\mstar}$, $k=2\pi s/V_{-}$, $\mstar = \kappa V_{-}/(2\pi)$ and $\gamma=V_{-}/V_{+}$. Then Eq. (\ref{eq:Consistency-Cond-inverse-1}) simplifies to the form
\begin{eqnarray}
\label{eq:Consistency-Cond-inverse-2}
\mathbb{S}_{-}(\kr) - \mathbb{S}_{+}(\kr) 
=\frac{\gamma}{4\pi^2 \mstar} \sum_{s=1}^{\infty} \frac{1}{s}\Big[ |\bar{z}_{-}(s)\mathcal{K}(1,\bar{z}_{-}(s))|^2 \nonumber\\
-  |\bar{z}_{+}\mathcal{K}(1,\bar{z}_{+}(s))|^2 \Big] .         
\end{eqnarray}
It is important to notice that Eq. (\ref{eq:Consistency-Cond-inverse-2}) is exactly similar to the 
second summation  term in Eq. (\ref{eq:Consistency-Cond-complete-rel-4}) (with $\alpha_{2}=0$ and 
$\alpha_{3}=1$), where we had neglected its contribution in the infinite volume limit, 
$\mstar\rightarrow\infty$. Here we shall analyse its behavior for small as well as large values of 
$s$ a bit more carefully.  For small values of $s$ \textit{i.e.,} when $\bar{z}_{\pm}\rightarrow 0$, 
$|\bar{z}_{\pm}K_{1}(\bar{z}_{\pm})|^2 \sim 1$ and the terms inside the square bracket cancels out 
exactly. Then only for large values of Bessel function's argument we get non zero contributions and 
that is
\begin{eqnarray}\label{eq:Consistency-Cond-inverse-3}
    \mathbb{S}_{-}(\kr) - \mathbb{S}_{+}(\kr) &=& \frac{\gamma \bar{a}}{8\pi \mstar}  \int_{s_{\star}}^{\infty} \frac{1}{\sqrt{s}}\Big[e^{-2\delta \bar{a} \sqrt{s}}- e^{-2 \bar{a} \sqrt{s}}\Big]ds \nonumber\\
    &=& \frac{\gamma}{8\pi\mstar}\Big[\frac{1}{\delta} e^{-2\delta \bar{a} \sqrt{s_{\star}}}- e^{-2 \bar{a} \sqrt{s_{\star}}}\Big]~,
\end{eqnarray}
where $\bar{a}=2\sqrt{|b_{0}|^2/\mstar}$. Here we have considered that in the range $s\in[s_{\star}, \infty)$, $\bar{z}_{\pm}\gg 1$. Then from Eq. (\ref{eq:Consistency-Cond-inverse-3}) the only dominating quantity, for small $\delta$, contributing in the consistency condition is
\begin{eqnarray}\label{eq:Consistency-Cond-inverse-4}
  \mathbb{S}_{-}(\kr) - \mathbb{S}_{+}(\kr) = \frac{1}{\delta}\frac{\gamma}{8\pi\mstar} =1~,
\end{eqnarray}
which was previously neglected in the $\mstar\to\infty$ limit considering a non-zero positive regularizing parameter $\delta$. However, in this case we cannot neglect it, otherwise the LHS of (\ref{eq:Consistency-Cond-inverse-4}) vanishes thus failing the consistency condition. One can further simplify this consistency condition to get, $4 \kr~\delta~ V_{+}=1$, which says that for a fixed frequency $\kr$ of the outgoing wave mode the integral regulator $\delta$ and the volume regulator $V_{+}$ are not independent and one is inversely proportional to the other. Then as $V_{+}\to \infty$ one should make the integral regulator $\delta\to 0$, which is in agreement with our initial assumptions on the regulators. We shall then keep this particular phenomena in mind in the estimation of the number density of the Hawking quanta in this case. Then we shall not make the quantities vanish when there is a $\gamma/(\delta\mstar)$ factor multiplied in any quantity.

\subsubsection{Number density}
Using the expression of Eq. (\ref{InvRelCoeff}) in Eq. (\ref{NumberDensityEVGeneral}) one can get the number density of Hawking quanta, now can be written as
\begin{eqnarray}\label{InvRelNumberDensity}
N_{\omegat} &=& \mathbb{S}_{+}(\kr) = \sum_{k>0} \left(\frac{k}{\kr}\right) |F_{1}^{\delta}(k,\kr)|^2 \nonumber\\
&=& \frac{\gamma}{4\pi^2 \mstar} \sum_{s=1}^{\infty} \frac{1}{s} |\bar{z}_{+}\mathcal{K}(1,\bar{z}_{+}(s))|^2 \nonumber\\
&\approx& \frac{\gamma}{4\pi^2 \mstar} \Bigg[\sum_{s=1}^{s_{L}}\frac{1}{s} + \sum_{s_{*}}^{\infty}\frac{\pi\bar{a}}{2\sqrt{s}}e^{-2\bar{a}\sqrt{s}}\Bigg],
\end{eqnarray}
where we have considered $\bar{z}_{+}\ll 1$ when $s\in[1,s_{L}]$ and $\bar{z}_{+}\gg 1$ for $s\in[s_{\star}, \infty)$. Here we observe that there is no $\delta$ in the denominator in the factor outside and the function inside the sum also do not contain $\delta$. Then we do not have any problem in taking the limit $\mstar\to \infty$ which makes this quantity to vanish. Another way to realize this is to use the consistency relation (\ref{eq:Consistency-Cond-inverse-4}) into (\ref{InvRelNumberDensity}) which follows
\begin{eqnarray}
N_{\omegat} = \frac{2}{\pi}\delta  \Bigg[\sum_{s=1}^{s_{L}}\frac{1}{s} + \sum_{s_{*}}^{\infty}\frac{\pi\bar{a}}{2\sqrt{s}}e^{-2\bar{a}\sqrt{s}}\Bigg]~.
\end{eqnarray}
In the limit $\delta\rightarrow 0$ the number density vanishes. This implies that in this situation also one can get a vanishing number density of Hawking quanta, i.e., $N_{\omegat} = \lim_{\mstar\to\infty} \mathbb{S}_{+}(\kr)=0$.

\section{Wien's displacement law and the understanding for a vanishing number density}\label{Sec:wiens-law}

In this part we intend to provide a description of a particle as it travels from $\scriminus$ to $\scriplus$ escaping getting trapped by the formation of the horizon. Through this description we expect to get an idea about the changes in a particle's characteristics due to the presence of the horizon as observed at the asymptotic future infinity. With this analysis we intend to provide a physical reasoning and identify the main contributing term behind the vanishing number density of the Hawking quanta in an extremal Kerr black hole spacetime. For the convenience of understanding we shall first consider the case of a non-extremal Kerr black hole and shall get into the extremal case subsequently. In a non-extremal Kerr black hole spacetime the relation among the spacelike near-null coordinates \cite{Barman:2018ina} is
\begin{equation}\label{eq:NN-rel-NE}
\xi_{+} = \xi_{-} + \frac{1}{\kappa_{h}} \ln{\left[\kappa_{h}\xi_{-}\right]}-\frac{1}{\kappa_{c}} \ln{\left[1+\frac{\kappa_{h}\xi_{-}}{\sigma}\right]}~,
\end{equation}
where, $\kappa_{h}$ and $\kappa_{c}$ are respectively the surface gravities at the event horizon and at the inner Cauchy horizon. Here, $\sigma=\kappa_{h}(\Delta_{c}-\Delta_{h})$, and $\Delta_{c}\equiv 2(r^0-r_{c})$, $\Delta_{h}\equiv 2(r^0-r_{h})$ with $r_{c}$, $r_{h}$ respective denoting the radius of the inner Cauchy horizon and the event horizon, and $r^{0}$ is a radial pivotal value considering which the geometric ray tracing is done to obtain the above relation among the spatial near-null coordinates. It should be mentioned that to obtain this relation the expression of the tortoise coordinate $\rstar= r + (1/2\kappa_{h}) \ln{\left[(r-r_{h})\kappa_{h}\right]} - (1/2\kappa_{c}) \ln{\left[(r-r_{c}) \kappa_{c}\right]}$, in a non-extremal Kerr black hole spacetime is utilized. On a constant time hypersurface using Eq. (\ref{eq:NN-rel-NE}) one can obtain the expression
\begin{equation}\label{eq:NN-rel-NE2}
\Delta\xi_{+} = \Delta\xi_{-} \left[ 1 + \frac{1}{\kappa_{h}\xi_{-}}-\frac{\frac{\kappa_{h}}{\kappa_{c}\sigma}}{1+\frac{\kappa_{h}\xi_{-}}{\sigma}} \right]~,
\end{equation}
where, $\Delta\xi_{+}$ can be identified to the De-Broglie wavelength $\lambda_{o}$ of the particle at future null infinity, and $\Delta\xi_{-}$ to the De-Broglie wavelength $\lambda_{e}$ of the particle at past null infinity. Then for a particle starting its journey from a spatial point $\xi^{e}_{-}$ the relation relating the wavelengths is
\begin{equation}\label{eq:NN-rel-NE3}
\lambda_{o} = \lambda_{e} \left[ 1 + \frac{1}{\kappa_{h}\xi^{e}_{-}}-\frac{\frac{\kappa_{h}}{\kappa_{c}\sigma}}{1+\kappa_{h}\xi^{e}_{-}/\sigma} \right]~.
\end{equation}
It is to be noted from (\ref{eq:NN-rel-NE}) that the modes responsible for the Planckian distribution of the Hawking effect are emitted from the region $\xi_{-}\ll 1/\kappa_{h}$ which results in the relation (\ref{eq:NN-rel-NE2}) to be $\xi_{+} \approx (1/\kappa_{h}) \ln{\left[\kappa_{h}\xi_{-}\right]}$. This phenomena can also be realized from Eq. (\ref{eq:NN-rel-NE3}) where it is noticed that for $\xi_{-}\gg 1/\kappa_{h}$ one has $\lambda_{o} \approx \lambda_{e}$, i.e., there is not much change in the characteristics of a particle that have passed long before the horizon formation. Then the particles that narrowly escape the formation of the event horizon can only contribute to the Planckian distribution of the Hawking effect. On the other hand, the point of emission $\xi^{e}_{-}$ cannot be made more accurate than the ingoing particle's De-Broglie wavelength $\lambda_{e}$, i.e., $\xi^{e}_{-}\approx\lambda_{e}$. Here again $\lambda_{e}$ can be expressed in terms of the ingoing mode's frequency $\lambda_{e}=hc/E^0_k=2hc/|k|$. Therefore, for the smallest possible wavelength of the particle one must consider the largest possible $|k|$, i.e., $|k|\to\infty$. Then in this limit it is seen that the wavelength of the particle observed at future null infinite is
\begin{equation}\label{eq:Wiens-law-NE}
\lambda_{o} \approx \frac{1}{\kappa_{h}}=\frac{2\pi}{T_{H}}~,
\end{equation}
which, signifies the Wien's displacement law for black body radiation, i.e., the characteristic temperature corresponding to a blackbody distribution is inversely proportional to the wavelength.

On the other hand, in an extremal Kerr black hole spacetime the relation among the spatial near-null coordinates is
\begin{equation}\label{eq:NN-rel-E}
\xi_{+} = \xi_{-} + 2r_{s} \ln{\left[\frac{\xi_{-}}{\sqrt{2}r_{s}}\right]}-\frac{2r_{s}^2}{\xi_{-}}~,
\end{equation}
where, the expression of the tortoise coordinate $\rstar= r + r_{s} \ln{\left[\xi_{-}/ (\sqrt{2}r_{s})\right]} - 2r_{s}^2/(2r-r_{s})$ is used to obtain this relation. From this expression one can obtain the relation among the observed and emitted wavelength of a particle, similar to Eq. (\ref{eq:NN-rel-NE3}), as
\begin{equation}\label{eq:NN-rel-E2}
\lambda_{o} = \lambda_{e} \left[ 1 + \frac{2r_{s}}{\xi^{e}_{-}} + \frac{2r_{s}^2}{\xi^{e^2}_{-}} \right]~.
\end{equation}
In this case if one adheres to the same concepts that the contribution significant to the Hawking effect comes from the high frequency modes nearly escaping the horizon formation, then it is convenient to choose $\xi^{e}_{-}\approx\lambda_{e}$ and then to take the limit $\lambda_{e}\to 0$. Unlike Eq. (\ref{eq:Wiens-law-NE}) in this case we observe that 
\begin{equation}\label{eq:Wiens-law-E}
\lambda_{o} \to \infty ~,
\end{equation}
which, corresponds to a characteristic temperature $T_{H}=0$ of the Hawking effect. It should be 
noted that the result of (\ref{eq:Wiens-law-E}) is due to the third quantity of the right hand side 
of Eq. (\ref{eq:NN-rel-E2}). Then it is evident that the inverse term in the relation between the 
spatial near-null coordinates in Eq. (\ref{eq:NN-rel-E}) is the dictating quantity. Presence of this 
inverse term in that relation results in a vanishing number density of the Hawking quanta for 
extremal Kerr black hole spacetime.

\section{Discussion}\label{Sec:discussion}

In this work, we have provided a detailed derivation of the vanishing number density of Hawking 
quanta in an extremal Kerr black hole spacetime using the canonical formulation 
\cite{Barman:2017fzh, Barman:2017vqx, Barman:2018ina}.
In \cite{Barman:2018ina} the authors used the canonical derivation to study the Hawking effect for 
both nonextremal and extremal Kerr black holes. However, the extremal case needed further studies to 
understand its origin of zero temperature better. In the derivation of the Hawking effect, the 
relation (\ref{eq:near-null-relation-extremal}) between the near-null coordinates near past 
and future null infinities plays a crucial role. In \cite{Barman:2018ina} an approximation was made 
on that relation for mathematical simplification. In the present work, we started with the full 
relation and consistently arrived at the zero temperature conclusion, solidifying the results of 
\cite{Barman:2018ina}. We also discussed the effects of different approximations of the relation on 
the final result and the consistency condition. Furthermore, we presented an argument to visualize 
the zero temperature from a physically understandable point of view, pinpointing the particular term 
in the relation mentioned above responsible for the phenomenon.

It is to be noted that the concept of the spacelike and timelike near-null coordinates, obtained by 
slightly deforming the null coordinates, is of vital importance to describe the dynamics of a matter 
field Hamiltonian in canonical formulation. Here we aimed to identify the term which contributes to 
the vanishing number density of the Hawking quanta in the relation between the spatial near-null 
coordinates. In this regard, we have first considered the entire relationship between the spatial 
near-null coordinates (\ref{eq:near-null-relation-extremal-f}) without any approximations and 
consistently obtained a vanishing number density of Hawking quanta. In Bogoluibov transformation 
method \cite{Gao:2002kz, Liberati:2000zz, Alvarenga:2003jd} one can obtain a similar relation like 
the inverse term from Eq. (\ref{eq:near-null-relation-extremal-f}) between the null coordinates. 
However, to the best of our knowledge, there is no detailed study with the complete relation. And 
this makes our study important and unique to a great extent. In this Hamiltonian formulation we 
first established the consistency condition which results from the simultaneous satisfaction of the 
Poisson brackets of the field modes and their conjugate momenta for the past and future observer. 
Using the relation (\ref{eq:near-null-relation-extremal-f}) to understand the Hawking effect, we 
have introduced a few parameters $\alpha_{j}$ (with $j=1,2,3$) which can have values $0$ or $1$ to 
keep track of the contributions of each term in the complete relation. However, we noticed from the 
final form of the consistency condition that one can not make $\alpha_{1}=0$ because we have already 
utilized a substitution containing $\alpha_{1}$ to arrive at the result and making it zero afterward 
will render the mathematics incorrect. We also observed that when $\alpha_{2}=0$ in Eq. 
(\ref{eq:near-null-relation-extremal-f}), i.e., the contribution of the logarithmic term is 
neglected, the consistency condition arrives to be the same as that was obtained in 
\cite{Barman:2018ina} and also the number density vanishes. Therefore from this first analysis, we 
conclude that the final result regarding the Hawking effect in an extremal Kerr black hole spacetime 
remains impervious to additional generalizations to the relation between spatial near-null 
coordinates as considered in \cite{Barman:2018ina}.

Furthermore, we have also considered only the inverse term in the relation 
(\ref{eq:near-null-relation-extremal-f}) for the study of the Hawking effect. In that case, we 
observed that for the domain $[0,\infty)$ of $\xi_{-}$ (which corresponds to the situation after the 
formation of the black hole event horizon), the other spatial near null coordinate $\xi_{+}$ does 
not cover the entire $(-\infty,\infty)$ instead covers a reduced domain $(-\infty,0]$. This 
observation has its physical shortcomings as it proclaims that not all future observers are eligible 
to comment about the Hawking effect (not even an observer at future timelike infinity). 
Nevertheless, mathematically one can pursue this case to obtain a vanishing number density of the 
Hawking quanta in a consistent manner. Our study solidifies the claims of a semi-classical approach 
in an extremal black hole spacetime.

This Hamiltonian-based formulation closely mimics the Bogoliubov transformation method of Hawking's 
original derivation \cite{hawking1975}. It opens the avenue to study the effects of other 
quantization techniques, like Polymer quantization \cite{Hossain:2010eb}, into the picture. 
Furthermore, from our current work, one can observe that it has a robust mathematical structure that 
can consistently answer some longstanding questions related to particle creation in extremal black 
hole backgrounds.

Finally, to provide a firm physical reasoning to recognize the main contributing term in the number 
density, we presented a thought experiment to realize the Hawking effect in a Kerr black hole 
spacetime. First, in a nonextremal Kerr black hole spacetime, we observed that the particles nearly 
escaping the formation of the event horizon and arriving at the scri-plus to be detected as the 
Hawking quanta must possess the final wavelength inversely proportional to the temperature of the 
Hawking effect. It establishes the \textit{Wien's displacement law} for thermal distribution of 
particles corresponding to the Hawking effect in a nonextremal Kerr black hole spacetime. Second, 
with the same setup in the extremal Kerr black hole spacetime, we found that the final wavelength of 
the particles approaches infinity. Compared to the nonextremal case, one can then associate the 
corresponding Hawking temperature to zero in the extremal scenario. We observed that this result is 
completely dictated by the inverse term in the relation (\ref{eq:near-null-relation-extremal-f}), 
thus providing us with a definitive understanding of the vanishing number density of the Hawking 
quanta.

Our analysis has presented many intricate findings regarding the Hawking effect in an extremal Kerr 
black hole spacetime. We believe that these results will further improve the understanding of 
particle creation in extremal black hole spacetimes.
It is to be noted that both Schwarzschild and the Kerr black hole spacetimes describe 
asymptotically flat geometry in regions far from the event horizon. A massless minimally coupled 
scalar field in these regions and near the horizon behaves like an infinite collection of fields 
from the flat spacetime, allowing one to construct the near-null coordinates and realize the Hawking 
effect using the Hamiltonian formulation \cite{Barman:2017fzh, Barman:2018ina}. In flat 
spacetime, one can realize the scalar field as an infinite sum of simple harmonic oscillators in the 
Fourier domain, which enables one to express the number density of Hawking quanta in terms of the 
Hamiltonian corresponding to each Fourier field mode. While one can successfully pursue this 
Hamiltonian-based formulation in flat spacetime \cite{Hossain:2014fma}, Schwarzschild, and Kerr 
black hole backgrounds, one cannot demand that the same will be true in other non-trivial black 
hole backgrounds. For example, in a de Sitter black hole background, the regions of radial infinity 
are not essentially flat, then one cannot readily apply the machinery of quantum field theory from 
flat spacetime in this background. But that remains an open direction to study further with the 
current approach.

\begin{acknowledgments}
The authors acknowledge Golam Mortuza Hossain and Bibhas Ranjan Majhi for useful discussions concerning the current topic. S.G. wishes to thank Indian Institute of Science Education and Research Kolkata (IISER Kolkata), and S.B. thanks Indian Institute of Technology Guwahati (IIT Guwahati) for financial support.
\end{acknowledgments}

\bibliographystyle{apsrev}

\bibliography{bibtexfile}

\begin{thebibliography}{48}
\expandafter\ifx\csname natexlab\endcsname\relax\def\natexlab#1{#1}\fi
\expandafter\ifx\csname bibnamefont\endcsname\relax
  \def\bibnamefont#1{#1}\fi
\expandafter\ifx\csname bibfnamefont\endcsname\relax
  \def\bibfnamefont#1{#1}\fi
\expandafter\ifx\csname citenamefont\endcsname\relax
  \def\citenamefont#1{#1}\fi
\expandafter\ifx\csname url\endcsname\relax
  \def\url#1{\texttt{#1}}\fi
\expandafter\ifx\csname urlprefix\endcsname\relax\def\urlprefix{URL }\fi
\providecommand{\bibinfo}[2]{#2}
\providecommand{\eprint}[2][]{\url{#2}}

\bibitem[{\citenamefont{Hawking}(1975)}]{hawking1975}
\bibinfo{author}{\bibfnamefont{S.~W.} \bibnamefont{Hawking}},
  \bibinfo{journal}{Comm. Math. Phys.} \textbf{\bibinfo{volume}{43}},
  \bibinfo{pages}{199} (\bibinfo{year}{1975}).

\bibitem[{\citenamefont{Barbachoux and Fabbri}(2002)}]{Barbachoux:2002tt}
\bibinfo{author}{\bibfnamefont{C.}~\bibnamefont{Barbachoux}} \bibnamefont{and}
  \bibinfo{author}{\bibfnamefont{A.}~\bibnamefont{Fabbri}},
  \bibinfo{journal}{Phys. Rev. D} \textbf{\bibinfo{volume}{66}},
  \bibinfo{pages}{024012} (\bibinfo{year}{2002}),
  \eprint{arXiv:hep-th/0201133}.

\bibitem[{\citenamefont{Angheben et~al.}(2005)\citenamefont{Angheben, Nadalini,
  Vanzo, and Zerbini}}]{Angheben:2005rm}
\bibinfo{author}{\bibfnamefont{M.}~\bibnamefont{Angheben}},
  \bibinfo{author}{\bibfnamefont{M.}~\bibnamefont{Nadalini}},
  \bibinfo{author}{\bibfnamefont{L.}~\bibnamefont{Vanzo}}, \bibnamefont{and}
  \bibinfo{author}{\bibfnamefont{S.}~\bibnamefont{Zerbini}},
  \bibinfo{journal}{JHEP} \textbf{\bibinfo{volume}{05}}, \bibinfo{pages}{014}
  (\bibinfo{year}{2005}), \eprint{arXiv:hep-th/0503081}.

\bibitem[{\citenamefont{Vanzo et~al.}(2011)\citenamefont{Vanzo, Acquaviva, and
  Di~Criscienzo}}]{Vanzo:2011wq}
\bibinfo{author}{\bibfnamefont{L.}~\bibnamefont{Vanzo}},
  \bibinfo{author}{\bibfnamefont{G.}~\bibnamefont{Acquaviva}},
  \bibnamefont{and}
  \bibinfo{author}{\bibfnamefont{R.}~\bibnamefont{Di~Criscienzo}},
  \bibinfo{journal}{Class. Quant. Grav.} \textbf{\bibinfo{volume}{28}},
  \bibinfo{pages}{183001} (\bibinfo{year}{2011}), \eprint{arXiv:1106.4153}.

\bibitem[{\citenamefont{Hayward et~al.}(2009)\citenamefont{Hayward,
  Di~Criscienzo, Vanzo, Nadalini, and Zerbini}}]{Hayward:2008jq}
\bibinfo{author}{\bibfnamefont{S.~A.} \bibnamefont{Hayward}},
  \bibinfo{author}{\bibfnamefont{R.}~\bibnamefont{Di~Criscienzo}},
  \bibinfo{author}{\bibfnamefont{L.}~\bibnamefont{Vanzo}},
  \bibinfo{author}{\bibfnamefont{M.}~\bibnamefont{Nadalini}}, \bibnamefont{and}
  \bibinfo{author}{\bibfnamefont{S.}~\bibnamefont{Zerbini}},
  \bibinfo{journal}{Class. Quant. Grav.} \textbf{\bibinfo{volume}{26}},
  \bibinfo{pages}{062001} (\bibinfo{year}{2009}), \eprint{arXiv:0806.0014}.

\bibitem[{\citenamefont{Anderson et~al.}(1995)\citenamefont{Anderson, Hiscock,
  and Loranz}}]{Anderson:1995fw}
\bibinfo{author}{\bibfnamefont{P.~R.} \bibnamefont{Anderson}},
  \bibinfo{author}{\bibfnamefont{W.~A.} \bibnamefont{Hiscock}},
  \bibnamefont{and} \bibinfo{author}{\bibfnamefont{D.~J.}
  \bibnamefont{Loranz}}, \bibinfo{journal}{Phys. Rev. Lett.}
  \textbf{\bibinfo{volume}{74}}, \bibinfo{pages}{4365} (\bibinfo{year}{1995}),
  \eprint{arXiv:gr-qc/9504019}.

\bibitem[{\citenamefont{Belgiorno}(1998)}]{BelgiornoF}
\bibinfo{author}{\bibfnamefont{F.}~\bibnamefont{Belgiorno}},
  \bibinfo{journal}{Journal of Mathematical Physics}
  \textbf{\bibinfo{volume}{39}}, \bibinfo{pages}{4608} (\bibinfo{year}{1998}).

\bibitem[{\citenamefont{Moretti}(1995)}]{Moretti:1995fu}
\bibinfo{author}{\bibfnamefont{V.}~\bibnamefont{Moretti}}
  (\bibinfo{year}{1995}), \eprint{arXiv:gr-qc/9510016}.

\bibitem[{\citenamefont{Eskin}(2019)}]{Eskin:2019nbo}
\bibinfo{author}{\bibfnamefont{G.}~\bibnamefont{Eskin}} (\bibinfo{year}{2019}),
  \eprint{arXiv:1902.05202}.

\bibitem[{\citenamefont{Parikh and Wilczek}(2000)}]{Parikh:1999mf}
\bibinfo{author}{\bibfnamefont{M.~K.} \bibnamefont{Parikh}} \bibnamefont{and}
  \bibinfo{author}{\bibfnamefont{F.}~\bibnamefont{Wilczek}},
  \bibinfo{journal}{Phys. Rev. Lett.} \textbf{\bibinfo{volume}{85}},
  \bibinfo{pages}{5042} (\bibinfo{year}{2000}), \eprint{hep-th/9907001}.

\bibitem[{\citenamefont{Gibbons and Hawking}(1977)}]{PhysRevD.15.2752}
\bibinfo{author}{\bibfnamefont{G.~W.} \bibnamefont{Gibbons}} \bibnamefont{and}
  \bibinfo{author}{\bibfnamefont{S.~W.} \bibnamefont{Hawking}},
  \bibinfo{journal}{Phys. Rev. D} \textbf{\bibinfo{volume}{15}},
  \bibinfo{pages}{2752} (\bibinfo{year}{1977}).

\bibitem[{\citenamefont{Alvarenga
  et~al.}(2003{\natexlab{a}})\citenamefont{Alvarenga, Batista, Fabris, and
  Marques}}]{Alvarenga:2003tx}
\bibinfo{author}{\bibfnamefont{F.~G.} \bibnamefont{Alvarenga}},
  \bibinfo{author}{\bibfnamefont{A.~B.} \bibnamefont{Batista}},
  \bibinfo{author}{\bibfnamefont{J.~C.} \bibnamefont{Fabris}},
  \bibnamefont{and} \bibinfo{author}{\bibfnamefont{G.~T.}
  \bibnamefont{Marques}}, \bibinfo{journal}{Phys. Lett.}
  \textbf{\bibinfo{volume}{A320}}, \bibinfo{pages}{83}
  (\bibinfo{year}{2003}{\natexlab{a}}), \eprint{arXiv:gr-qc/0306030}.

\bibitem[{\citenamefont{Barman et~al.}(2018)\citenamefont{Barman, Hossain, and
  Singha}}]{Barman:2017fzh}
\bibinfo{author}{\bibfnamefont{S.}~\bibnamefont{Barman}},
  \bibinfo{author}{\bibfnamefont{G.~M.} \bibnamefont{Hossain}},
  \bibnamefont{and} \bibinfo{author}{\bibfnamefont{C.}~\bibnamefont{Singha}},
  \bibinfo{journal}{Phys. Rev.} \textbf{\bibinfo{volume}{D97}},
  \bibinfo{pages}{025016} (\bibinfo{year}{2018}), \eprint{arXiv:1707.03614}.

\bibitem[{\citenamefont{Barman et~al.}(2019)\citenamefont{Barman, Hossain, and
  Singha}}]{Barman:2017vqx}
\bibinfo{author}{\bibfnamefont{S.}~\bibnamefont{Barman}},
  \bibinfo{author}{\bibfnamefont{G.~M.} \bibnamefont{Hossain}},
  \bibnamefont{and} \bibinfo{author}{\bibfnamefont{C.}~\bibnamefont{Singha}},
  \bibinfo{journal}{J. Math. Phys.} \textbf{\bibinfo{volume}{60}},
  \bibinfo{pages}{052304} (\bibinfo{year}{2019}), \eprint{arXiv:1707.03605}.

\bibitem[{\citenamefont{Barman and Hossain}(2019)}]{Barman:2018ina}
\bibinfo{author}{\bibfnamefont{S.}~\bibnamefont{Barman}} \bibnamefont{and}
  \bibinfo{author}{\bibfnamefont{G.~M.} \bibnamefont{Hossain}},
  \bibinfo{journal}{Phys. Rev. D} \textbf{\bibinfo{volume}{99}},
  \bibinfo{pages}{065010} (\bibinfo{year}{2019}), \eprint{arXiv:1809.09430}.

\bibitem[{\citenamefont{Liberati et~al.}(2000)\citenamefont{Liberati, Rothman,
  and Sonego}}]{Liberati:2000sq}
\bibinfo{author}{\bibfnamefont{S.}~\bibnamefont{Liberati}},
  \bibinfo{author}{\bibfnamefont{T.}~\bibnamefont{Rothman}}, \bibnamefont{and}
  \bibinfo{author}{\bibfnamefont{S.}~\bibnamefont{Sonego}},
  \bibinfo{journal}{Phys. Rev.} \textbf{\bibinfo{volume}{D62}},
  \bibinfo{pages}{024005} (\bibinfo{year}{2000}), \eprint{arXiv:gr-qc/0002019}.

\bibitem[{\citenamefont{Abbott et~al.}(2016{\natexlab{a}})}]{Abbott:2016blz}
\bibinfo{author}{\bibfnamefont{B.~P.} \bibnamefont{Abbott}}
  \bibnamefont{et~al.} (\bibinfo{collaboration}{Virgo, LIGO Scientific}),
  \bibinfo{journal}{Phys. Rev. Lett.} \textbf{\bibinfo{volume}{116}},
  \bibinfo{pages}{061102} (\bibinfo{year}{2016}{\natexlab{a}}),
  \eprint{arXiv:1602.03837}.

\bibitem[{\citenamefont{Abbott et~al.}(2016{\natexlab{b}})}]{Abbott:2016nmj}
\bibinfo{author}{\bibfnamefont{B.~P.} \bibnamefont{Abbott}}
  \bibnamefont{et~al.} (\bibinfo{collaboration}{Virgo, LIGO Scientific}),
  \bibinfo{journal}{Phys. Rev. Lett.} \textbf{\bibinfo{volume}{116}},
  \bibinfo{pages}{241103} (\bibinfo{year}{2016}{\natexlab{b}}),
  \eprint{arXiv:1606.04855}.

\bibitem[{\citenamefont{Abbott
  et~al.}(2016{\natexlab{c}})}]{TheLIGOScientific:2016pea}
\bibinfo{author}{\bibfnamefont{B.~P.} \bibnamefont{Abbott}}
  \bibnamefont{et~al.} (\bibinfo{collaboration}{Virgo, LIGO Scientific}),
  \bibinfo{journal}{Phys. Rev.} \textbf{\bibinfo{volume}{X6}},
  \bibinfo{pages}{041015} (\bibinfo{year}{2016}{\natexlab{c}}),
  \eprint{arXiv:1606.04856}.

\bibitem[{\citenamefont{Abbott et~al.}(2017)}]{Abbott:2017vtc}
\bibinfo{author}{\bibfnamefont{B.~P.} \bibnamefont{Abbott}}
  \bibnamefont{et~al.} (\bibinfo{collaboration}{VIRGO, LIGO Scientific}),
  \bibinfo{journal}{Phys. Rev. Lett.} \textbf{\bibinfo{volume}{118}},
  \bibinfo{pages}{221101} (\bibinfo{year}{2017}), \eprint{arXiv:1706.01812}.

\bibitem[{\citenamefont{{Boyer} and {Lindquist}}(1967)}]{Boyer:1967}
\bibinfo{author}{\bibfnamefont{R.~H.} \bibnamefont{{Boyer}}} \bibnamefont{and}
  \bibinfo{author}{\bibfnamefont{R.~W.} \bibnamefont{{Lindquist}}},
  \bibinfo{journal}{Journal of Mathematical Physics}
  \textbf{\bibinfo{volume}{8}}, \bibinfo{pages}{265} (\bibinfo{year}{1967}).

\bibitem[{\citenamefont{Carroll}(2004)}]{BookCarroll}
\bibinfo{author}{\bibfnamefont{S.}~\bibnamefont{Carroll}},
  \emph{\bibinfo{title}{Spacetime and geometry. An introduction to general
  relativity}} (\bibinfo{publisher}{AW}, \bibinfo{year}{2004}).

\bibitem[{\citenamefont{Kerr}(2007)}]{Kerr2007dk}
\bibinfo{author}{\bibfnamefont{R.~P.} \bibnamefont{Kerr}}, in
  \emph{\bibinfo{booktitle}{{Kerr Fest: Black Holes in Astrophysics, General
  Relativity and Quantum Gravity Christchurch, New Zealand, August 26-28,
  2004}}} (\bibinfo{year}{2007}), \eprint{arXiv:0706.1109}.

\bibitem[{\citenamefont{Poisson}(2007)}]{BookPoisson}
\bibinfo{author}{\bibfnamefont{E.}~\bibnamefont{Poisson}},
  \emph{\bibinfo{title}{A Relativist's Toolkit: The Mathematics of Black-Hole
  Mechanics}} (\bibinfo{publisher}{Cambridge University Press},
  \bibinfo{year}{2007}).

\bibitem[{\citenamefont{Dadhich}(2013)}]{Dadhich2013qx}
\bibinfo{author}{\bibfnamefont{N.}~\bibnamefont{Dadhich}},
  \bibinfo{journal}{Gen. Rel. Grav.} \textbf{\bibinfo{volume}{45}},
  \bibinfo{pages}{2383} (\bibinfo{year}{2013}), \eprint{arXiv:1301.5314}.

\bibitem[{\citenamefont{Schutz}(1985)}]{BookSchutz}
\bibinfo{author}{\bibfnamefont{B.~F.} \bibnamefont{Schutz}},
  \emph{\bibinfo{title}{A first course in general relativity}}
  (\bibinfo{publisher}{Cambridge University Press}, \bibinfo{year}{1985}).

\bibitem[{\citenamefont{Raine and Thomas}(2005)}]{BookRaine}
\bibinfo{author}{\bibfnamefont{D.}~\bibnamefont{Raine}} \bibnamefont{and}
  \bibinfo{author}{\bibfnamefont{E.}~\bibnamefont{Thomas}},
  \emph{\bibinfo{title}{Black Holes: An Introduction}}
  (\bibinfo{publisher}{Imperial College Press}, \bibinfo{year}{2005}).

\bibitem[{\citenamefont{Padmanabhan}(2010)}]{BookPadmanabhanGrav}
\bibinfo{author}{\bibfnamefont{T.}~\bibnamefont{Padmanabhan}},
  \emph{\bibinfo{title}{Gravitation: Foundations and Frontiers}}
  (\bibinfo{publisher}{Cambridge University Press}, \bibinfo{year}{2010}),
  \bibinfo{edition}{1st} ed.

\bibitem[{\citenamefont{Heinicke and Hehl}(2014)}]{Heinicke2015iva}
\bibinfo{author}{\bibfnamefont{C.}~\bibnamefont{Heinicke}} \bibnamefont{and}
  \bibinfo{author}{\bibfnamefont{F.~W.} \bibnamefont{Hehl}},
  \bibinfo{journal}{Int. J. Mod. Phys.} \textbf{\bibinfo{volume}{D24}},
  \bibinfo{pages}{1530006} (\bibinfo{year}{2014}), \eprint{arXiv:1503.02172}.

\bibitem[{\citenamefont{Krasinski}(1978)}]{Krasinski1900zzb}
\bibinfo{author}{\bibfnamefont{A.}~\bibnamefont{Krasinski}},
  \bibinfo{journal}{Annals Phys.} \textbf{\bibinfo{volume}{112}},
  \bibinfo{pages}{22} (\bibinfo{year}{1978}).

\bibitem[{\citenamefont{Teukolsky}(2015)}]{Teukolsky2014vca}
\bibinfo{author}{\bibfnamefont{S.~A.} \bibnamefont{Teukolsky}},
  \bibinfo{journal}{Class. Quant. Grav.} \textbf{\bibinfo{volume}{32}},
  \bibinfo{pages}{124006} (\bibinfo{year}{2015}), \eprint{arXiv:1410.2130}.

\bibitem[{\citenamefont{Visser}(2007)}]{Visser2007fj}
\bibinfo{author}{\bibfnamefont{M.}~\bibnamefont{Visser}}, in
  \emph{\bibinfo{booktitle}{{Kerr Fest: Black Holes in Astrophysics, General
  Relativity and Quantum Gravity Christchurch, New Zealand, August 26-28,
  2004}}} (\bibinfo{year}{2007}), \eprint{arXiv:0706.0622}.

\bibitem[{\citenamefont{Smailagic and Spallucci}(2010)}]{Smailagic2010nv}
\bibinfo{author}{\bibfnamefont{A.}~\bibnamefont{Smailagic}} \bibnamefont{and}
  \bibinfo{author}{\bibfnamefont{E.}~\bibnamefont{Spallucci}},
  \bibinfo{journal}{Phys. Lett.} \textbf{\bibinfo{volume}{B688}},
  \bibinfo{pages}{82} (\bibinfo{year}{2010}), \eprint{arXiv:1003.3918}.

\bibitem[{\citenamefont{Gao}(2003)}]{Gao:2002kz}
\bibinfo{author}{\bibfnamefont{S.}~\bibnamefont{Gao}}, \bibinfo{journal}{Phys.
  Rev.} \textbf{\bibinfo{volume}{D68}}, \bibinfo{pages}{044028}
  (\bibinfo{year}{2003}), \eprint{arXiv:gr-qc/0207029}.

\bibitem[{\citenamefont{Iso et~al.}(2006)\citenamefont{Iso, Umetsu, and
  Wilczek}}]{Iso:2006ut}
\bibinfo{author}{\bibfnamefont{S.}~\bibnamefont{Iso}},
  \bibinfo{author}{\bibfnamefont{H.}~\bibnamefont{Umetsu}}, \bibnamefont{and}
  \bibinfo{author}{\bibfnamefont{F.}~\bibnamefont{Wilczek}},
  \bibinfo{journal}{Phys. Rev.} \textbf{\bibinfo{volume}{D74}},
  \bibinfo{pages}{044017} (\bibinfo{year}{2006}), \eprint{hep-th/0606018}.

\bibitem[{\citenamefont{Ford}(1975)}]{Ford:1975tp}
\bibinfo{author}{\bibfnamefont{L.~H.} \bibnamefont{Ford}},
  \bibinfo{journal}{Phys. Rev.} \textbf{\bibinfo{volume}{D12}},
  \bibinfo{pages}{2963} (\bibinfo{year}{1975}).

\bibitem[{\citenamefont{Hod}(2011)}]{Hod:2011zzd}
\bibinfo{author}{\bibfnamefont{S.}~\bibnamefont{Hod}}, \bibinfo{journal}{Phys.
  Rev.} \textbf{\bibinfo{volume}{D84}}, \bibinfo{pages}{044046}
  (\bibinfo{year}{2011}), \eprint{arXiv:1109.4080}.

\bibitem[{\citenamefont{Menezes}(2017)}]{Menezes:2016SeaKs}
\bibinfo{author}{\bibfnamefont{G.}~\bibnamefont{Menezes}},
  \bibinfo{journal}{Phys. Rev.} \textbf{\bibinfo{volume}{D95}},
  \bibinfo{pages}{065015} (\bibinfo{year}{2017}), \eprint{arXiv:1611.00056}.

\bibitem[{\citenamefont{Menezes}(2018)}]{Menezes:2017oeb}
\bibinfo{author}{\bibfnamefont{G.}~\bibnamefont{Menezes}},
  \bibinfo{journal}{Phys. Rev.} \textbf{\bibinfo{volume}{D97}},
  \bibinfo{pages}{085021} (\bibinfo{year}{2018}), \eprint{arXiv:1712.07151}.

\bibitem[{\citenamefont{Lin and Soo}(2013)}]{Lin:2009wm}
\bibinfo{author}{\bibfnamefont{C.-Y.} \bibnamefont{Lin}} \bibnamefont{and}
  \bibinfo{author}{\bibfnamefont{C.}~\bibnamefont{Soo}}, \bibinfo{journal}{Gen.
  Rel. Grav.} \textbf{\bibinfo{volume}{45}}, \bibinfo{pages}{79}
  (\bibinfo{year}{2013}), \eprint{arXiv:0905.3244}.

\bibitem[{\citenamefont{Miao et~al.}(2012)\citenamefont{Miao, Xue, and
  Zhang}}]{Miao:2011dy}
\bibinfo{author}{\bibfnamefont{Y.-G.} \bibnamefont{Miao}},
  \bibinfo{author}{\bibfnamefont{Z.}~\bibnamefont{Xue}}, \bibnamefont{and}
  \bibinfo{author}{\bibfnamefont{S.-J.} \bibnamefont{Zhang}},
  \bibinfo{journal}{Int. J. Mod. Phys.} \textbf{\bibinfo{volume}{D21}},
  \bibinfo{pages}{1250018} (\bibinfo{year}{2012}), \eprint{arXiv:1102.0074}.

\bibitem[{\citenamefont{Vanzo}(1997)}]{Vanzo:1995bh}
\bibinfo{author}{\bibfnamefont{L.}~\bibnamefont{Vanzo}},
  \bibinfo{journal}{Phys. Rev.} \textbf{\bibinfo{volume}{D55}},
  \bibinfo{pages}{2192} (\bibinfo{year}{1997}), \eprint{arXiv:gr-qc/9510011}.

\bibitem[{\citenamefont{Balbinot et~al.}(2007)\citenamefont{Balbinot, Fabbri,
  Farese, and Parentani}}]{Balbinot:2007kr}
\bibinfo{author}{\bibfnamefont{R.}~\bibnamefont{Balbinot}},
  \bibinfo{author}{\bibfnamefont{A.}~\bibnamefont{Fabbri}},
  \bibinfo{author}{\bibfnamefont{S.}~\bibnamefont{Farese}}, \bibnamefont{and}
  \bibinfo{author}{\bibfnamefont{R.}~\bibnamefont{Parentani}},
  \bibinfo{journal}{Phys. Rev.} \textbf{\bibinfo{volume}{D76}},
  \bibinfo{pages}{124010} (\bibinfo{year}{2007}), \eprint{arXiv:0710.0388}.

\bibitem[{\citenamefont{Hossain et~al.}(2010)\citenamefont{Hossain, Husain, and
  Seahra}}]{Hossain:2010eb}
\bibinfo{author}{\bibfnamefont{G.~M.} \bibnamefont{Hossain}},
  \bibinfo{author}{\bibfnamefont{V.}~\bibnamefont{Husain}}, \bibnamefont{and}
  \bibinfo{author}{\bibfnamefont{S.~S.} \bibnamefont{Seahra}},
  \bibinfo{journal}{Phys.Rev.} \textbf{\bibinfo{volume}{D82}},
  \bibinfo{pages}{124032} (\bibinfo{year}{2010}), \eprint{arXiv:1007.5500}.

\bibitem[{{\relax DLMF}()}]{NIST:DLMF}
{\relax DLMF}, \emph{\bibinfo{title}{{\it NIST Digital Library of Mathematical
  Functions}}}, \bibinfo{note}{f.~W.~J. Olver, A.~B. {Olde Daalhuis}, D.~W.
  Lozier, B.~I. Schneider, R.~F. Boisvert, C.~W. Clark, B.~R. Miller and B.~V.
  Saunders, eds.}

\bibitem[{\citenamefont{Liberati et~al.}(2001)\citenamefont{Liberati, Rothman,
  and Sonego}}]{Liberati:2000zz}
\bibinfo{author}{\bibfnamefont{S.}~\bibnamefont{Liberati}},
  \bibinfo{author}{\bibfnamefont{T.}~\bibnamefont{Rothman}}, \bibnamefont{and}
  \bibinfo{author}{\bibfnamefont{S.}~\bibnamefont{Sonego}},
  \bibinfo{journal}{Int. J. Mod. Phys.} \textbf{\bibinfo{volume}{D10}},
  \bibinfo{pages}{33} (\bibinfo{year}{2001}), \eprint{arXiv:gr-qc/0008018}.

\bibitem[{\citenamefont{Alvarenga
  et~al.}(2003{\natexlab{b}})\citenamefont{Alvarenga, Batista, Fabris, and
  Marques}}]{Alvarenga:2003jd}
\bibinfo{author}{\bibfnamefont{F.~G.} \bibnamefont{Alvarenga}},
  \bibinfo{author}{\bibfnamefont{A.~B.} \bibnamefont{Batista}},
  \bibinfo{author}{\bibfnamefont{J.~C.} \bibnamefont{Fabris}},
  \bibnamefont{and} \bibinfo{author}{\bibfnamefont{G.~T.}
  \bibnamefont{Marques}} (\bibinfo{year}{2003}{\natexlab{b}}),
  \eprint{arXiv:gr-qc/0309022}.

\bibitem[{\citenamefont{Hossain and Sardar}(2016)}]{Hossain:2014fma}
\bibinfo{author}{\bibfnamefont{G.~M.} \bibnamefont{Hossain}} \bibnamefont{and}
  \bibinfo{author}{\bibfnamefont{G.}~\bibnamefont{Sardar}},
  \bibinfo{journal}{Class. Quant. Grav.} \textbf{\bibinfo{volume}{33}},
  \bibinfo{pages}{245016} (\bibinfo{year}{2016}), \eprint{arXiv:1411.1935}.

\end{thebibliography}

\end{document}